\documentclass[10pt,journal,final,twocolumn]{IEEEtran}
\IEEEoverridecommandlockouts
\usepackage{cite}
\usepackage{amsmath,amssymb,amsfonts}
\usepackage{algorithmic}
\usepackage{graphicx}
\usepackage{textcomp}
\usepackage{xcolor}
\usepackage{mathrsfs}
\usepackage{algorithm}
\usepackage{epstopdf}
\usepackage{psfrag}    
\usepackage{subfigure} 
\usepackage{url}       
\usepackage{stfloats}  
\usepackage{enumerate}
\usepackage{bm}
\usepackage{fancyhdr}
\usepackage{threeparttable}
\usepackage{setspace}
\usepackage{booktabs}
\usepackage{float}
\usepackage{graphicx,adjustbox,array}
\usepackage{caption}
\usepackage{diagbox}
\usepackage{hyperref}
\allowdisplaybreaks[4]
\hypersetup{hypertex=true,
            colorlinks=true,
            linkcolor=blue,
            anchorcolor=blue,
            citecolor=blue}
\makeatletter
\renewcommand{\maketag@@@}[1]{\hbox{\m@th\normalsize\normalfont#1}}
\makeatother

\def\BibTeX{{\rm B\kern-.05em{\sc i\kern-.025em b}\kern-.08em
    T\kern-.1667em\lower.7ex\hbox{E}\kern-.125emX}}

\begin{document}

\title{Sensing, Detection and Localization for Low Altitude UAV: A RF-Based Framework via Multiple BSs Collaboration}

\author{Tianhao Liang, \IEEEmembership{Member,~IEEE}, Mu Jia, Tingting Zhang, \IEEEmembership{Member,~IEEE}, Junting Chen, \IEEEmembership{Member,~IEEE}, Longyu Zhou, \IEEEmembership{Member,~IEEE}, Tony Q. S. Quek, \IEEEmembership{Fellow,~IEEE}, and Pooi-Yuen Kam, \IEEEmembership{Life Fellow,~IEEE}

\thanks{
Tianhao Liang and Tingting Zhang are with Guangdong Provincial Key Laboratory of Aerospace Communication and Networking Technology, School of Information Science and Technology, Harbin Institute of Technology (Shenzhen), Shenzhen, P.R. China. Tingting Zhang is also with Pengcheng Laboratory (PCL), Shenzhen, P.R. China. Mu Jia, Junting Chen, Pooi-Yuen Kam are with the School of Science and Engineering, Shenzhen Future Network of Intelligence Institute (FNii-Shenzhen), and Guangdong Provincial Key Laboratory of Future Networks of Intelligence, The Chinese University of Hong Kong, Shenzhen, Guangdong 518172, P.R. China. Longyu Zhou is with Information System Technology and Design, Singapore University of Technology and Design and also with ChinaTelecom Singapore Innovation Research Institute. Tony Q. S. Quek is with Information System Technology and Design Pillar, Singapore University of Technology and Design, Singapore.  (e-mail: liangth@hit.edu.cn; mujia1@link.cuhk.edu.cn; zhangtt@hit.edu.cn; juntingc@cuhk.edu.cn; zhoulyfuture@outlook.com; tonyquek@sutd.edu.sg; pykam@cuhk.edu.cn).
}
}

\maketitle

\begin{abstract}
The rapid growth of the low-altitude economy has resulted in a significant increase in the number of Low, slow, and small (LLS) unmanned aerial vehicles (UAVs), raising critical challenges for secure airspace management and reliable trajectory planning. To address this, this paper proposes a cooperative radio-frequency (RF) detection and localization framework that leverages existing cellular base stations. The proposed approach features a robust scheme for LSS target identification, integrating a cell averaging-constant false alarm rate (CA-CFAR) detector with a micro-Doppler signature (MDS) based recognition method.  Multi-station measurements are fused through a grid-based probabilistic algorithm combined with clustering techniques, effectively mitigating ghost targets and improving localization accuracy in multi-UAV scenarios. Furthermore, the Cramer-Rao lower bound (CRLB) is derived as a performance benchmark and reinforcement learning (RL)-based optimization is employed to balance localization accuracy against station resource usage. Simulations demonstrate that increasing from one to multiple BSs reduces the positioning error to near the CRLB, while practical experiments further verify the framework's effectiveness. Furthermore, our RL-based optimization can find solutions that maintain high accuracy while minimizing resource usage, highlighting its potential as a scalable solution for ensuring airspace safety in the emerging low-altitude economy.
\end{abstract}

\begin{IEEEkeywords}
Low altitude economy, UAV detection, cooperative localization, data fusion, Cramer-Rao bound.
\end{IEEEkeywords}

\section{Introduction}
\subsection{Background and Motivation}
Driven by the advancements in Unmanned aerial vehicle (UAV) technologies and supportive governmental policies, the low-altitude economy is experiencing rapid expansion from ground into the airspace. This growing promotes a wide range of emerging aerial applications, including logistics, emergency rescue, situation awareness, and environmental monitoring \cite{huang2024potential,he2025ubiquitous,ping2025multimodal,wu2021comprehensive,LTHNETWORK,lth2022icc,lth2024twc}.  Concurrently, this trend is leading to an unprecedented increase in the number of UAVs, which introduces complex challenges to airspace management and flight safety \cite{pothana2023uas, yang2024access,yuan2020wcl}.

On the one hand, diverse low-altitude services involve heterogeneous UAV types and mission requirements \cite{jin2025advancing}. In the absence of interconnectivity among these applications, UAVs from different services can act as non-cooperative aerial obstacles. Their trajectories need to be accurately sensed, detected and localized to enable collision avoidance and improve flight efficiency \cite{wang2025twc}, \cite{wu2025toward}. On the other hand, the extensive  use of UAVs may raise the risk of unauthorized and non-cooperative drones operations, posing potential threats to public safety, such as terrorist attack, smuggling, and the disruptions to the normal operations of critical infrastructure like airports, military strongholds and large hydroelectric plants \cite{cohen2021urban,dong2025iotj,huang2021icca}. Consequently, there is a pressing need for robust and reliable UAV sensing, detection and localization technologies, which are not only vital to ensure secure and reliable low-altitude operations, but also essential for supporting the sustainable development of the low-altitude economy \cite{lei2025enhancing,li2025iotj,liu2025mape,lth2024jsas,lth2025iotj}.

To achieve accurate detection accuracy, vision-based method have been investigated as mainstream UAV detection schemes \cite{niu2021icba}.  However, their performance degrades in occluded and low-light environments. In contrast, radio frequency (RF)-based methods can work all-time and all-weather, making them priority alternatives for robust UAV detection \cite{wang2024spl}, \cite{ezuma2020ojcs}. Generally, achieving high sensing, detection and localization accuracy in practice requires the deployment of numerous dedicated devices, which significantly increases infrastructure costs and prohibit their adoption in a short time \cite{xie2024iotj}. Prompted by the advancement of millimeter wave and integrated sensing and communication (ISAC) technologies for the upcoming sixth-generation (6G) networks, sensing functions can be performed directly by base stations (BSs) without requiring additional equipments \cite{yuan2021integrated,JiaYanZha:C22,liu2020radar,liu2022integrated,haonan2023pimrc,zhang2024predictive}. Inspired by investigations of the ISAC applications for aerial scenarios, UAV sensing, detection and localization can be implemented using the current cellular networks, offering a cost-effective and scalable solution by unifying spectral resources and hardware platforms \cite{yan2025deep,wu2025maga,wu2023uavs}.

Despite the growing attention to UAV sensing, detection and localization via BSs, practical operations still faces several key limitations \cite{zhao2023tgrs}, \cite{zhu2022tpami}, presented below. 
\begin{itemize}
  \item {UAVs in low-altitude applications are typically small targets, with a radar cross section (RCS) under 2$\rm{m^2}$, and their speeds are usually not exceeding 200$\rm{km/h}$. These characteristics classify them as low-slow-small (LSS) aerial targets, which are inherently difficult to distinguish from background clutter in urban environments, as they exhibit a low signal-to-clutter-plus-noise ratio and weak Doppler signatures \cite{wang2025sensor,selvi2025wisp,yuan2025sensor}.}
  \item {In highly dynamic low-altitude environments, many existing solutions exhibit excessive computational complexity, leading them difficult to satisfy real-time sensing, detection and localization requirements \cite{ahir2024map}, \cite{ak2024tvt}.}
  \item {Most importantly, the complex propagation environments in urban and low-altitude settings create significant multipath reflections. These reflections can generate false observations, named ``ghost'' targets, which severely degrade the accuracy and reliability of target detection \cite{GaoJiaZha:C21,ghostgeometrypaper,wanJiaMen:C22,ghostdataset}.}
\end{itemize}
These limitations highlight the urgent exploration for more advanced sensing, detection and localization frameworks of LLS targets within existing networks to meet the safety and reliability requirements of the emerging low-altitude economy.

Motivated by these challenges, this paper focuses on reliable UAV detection and localization by leveraging existing BSs, thereby reducing the need for costly dedicated sensing infrastructure.  Utilizing the antenna array at the BS, the hybrid time of arrival (ToA) and angle of arrival (AoA) method can be employed to obtain ranging and angle information from the UAV using pilot or ISAC signals \cite{ghoz2021wim}, \cite{jiayan2022iotj}. To enhance the detection and localization performance of BSs, the inherent Doppler modulation caused by the rotation of UAV's rotor, named micro-Doppler signature (MDS), can be exploited to obtain the more concentrated energy of LLS UAV targets \cite{jin2021access} \cite{sun2021tgrs}.
Furthermore, the cooperative scheme among multiple BSs and the efficient methods for processing multiple UAV targets recognition are necessary to be designed to eliminate ghost interference and enhance localization accuracy \cite{ztt2016twc}, which is the focus of this investigation.

\subsection{Related Works}
UAV detection and localization by multiple stations can be briefly divided into three key phases \cite{Bek2006TSP}. The first is parameter estimation process, where localization-related information (e.g., distance and angle) is extracted from the received RF signals. The second phase involves clutter suppression and feature extraction operations, which are critical for mitigating interference from the environment and distinguishing UAV targets from noise and ghost reflections. The last stage is the cooperative data fusion process, where information from multiple stations is combined to refine the final position of the UAV and improve overall accuracy \cite{yuan2016tvt}. In the following, we overview the related works corresponding to each of these three stages.

For UAV detection and localization, the estimation process of UAV parameters is determined by the localization method. Common methods include ToA, AoA, received signal strength (RSS), and their combinations \cite{ref6}.  Generally, RSS is usually applied in indoor localization systems due to its feasibility and simplicity. However, it exhibits low accuracy in environments with significant fading or unknown pathloss coefficient \cite{zafar2019rss,jia2025optimum}. Conversely, ToA and AoA methods can provide relative higher accuracy distance measurements, especially for large bandwidth signals. Due to the synchronization issues, the ToA measurements can be further categorized into time difference of arrival (TDOA) and round trip measuring methods \cite{wang2020tcom}.   Different from traditional distributed multiple input multiple output (MIMO) radar system \cite{MIMOradar},  the asynchronous problem has been effectively addressed in currently BS architectures. To fully utilize the capabilities  of these BSs, the hybrid TOA and AOA is often adopted to locate the detected targets, which has been proven to yield excellent localization performance \cite{lth2021vtc, lth2023vtc, yjy2021vtc}.

Clutters caused by the scatterers are the useless components in the received signals and must be suppressed. The key of clutter suppression is analyzing the property of clutter according to the relationship between scatterers and receiver systems \cite{clutterproperty}. For static reflectors and slow-moving clutter, benefited from the increased stability of digital processer, the moving target indicator (MTI) filters are effective at isolating targets with significant velocity \cite{MTIbasic}. However, for the dynamic LLS UAV targets with weak Doppler signatures, it is difficult to obtain high detection performance through the traditional detectors like constant false alarm rate (CFAR) \cite{ezu2019radar} and adaptive normalized matched filter (AMF) \cite{kam2018taes}. To overcome this issue, method based on the MDS have been extensively investigated.  Typically, the echo of rotor UAV mainly includes the translational component generated by the motion of UAV's body and the micro-Doppler component generated by the rotation of rotor, whose echo is a typical sine frequency modulation (SFM) signal \cite{rahman2018radar}. Therefore, by properly exploiting the micro-Doppler effects, target recognition can be significantly improved \cite{RadarConfMicro}. Generally, time-frequency analysis methods, such as short-time Fourier transform (STFT), radon-Wigner transform (RWT), etc., are used to generate the time-frequency spectrogram  representing the MDS \cite{fu2018cm,JiaShaKer:C20}.  Features are then extracted from these spectrograms using transition methods like singular value decomposition (SVD) \cite{vov2025tap}, empirical mode decomposition (EMD) \cite{EMD1}. Subsequently, the LLS UAV can be detected using classification methods, such as support vector machines (SVM), and convolutional neural networks (CNN) \cite{yu2025tifs}, \cite{khan2024tvt}.

The multipath propagation generated by random reflectors or unexpected obstacles is another critical issue during detection process, especially in the complicated environment \cite{Liu2024ics}. Compared to traditional long range detection, sensing inside the low-altitude scenarios suffers much more from the multipath effects, which leads to severe detection performance degradation and ghost targets generation \cite{zhou2021vtc}. Without proper priori knowledge of the environment, these ghosts could be mapped back to their corresponding true target location, which cause severe interference to the environmental sensing and mapping. Typically, two different ghost targets can be generated both from measurement procedure and data fusion phases, respectively. There may exist multipath ghosts produced by the indirect path propagations for an individual transmitter, with different bounce process from the reflector \cite{ghostbackground}. Additionally, multiple targets inside the beam intersection areas from multiple transmitter can also generate ghost detections \cite{beamintersection}. These ghost targets will lead to unexpected false alarms, which has attracted wide interests \cite{bistaticghost,ghostfusion}. Consequently, various techniques for ghost suppression have been investigated. In \cite{ghostbackground}, based on the phase property of multipath ghost, the phase coherent factor (PCF) was weighted to the radar image, to mitigate the ghost power. The authors in \cite{ghostrangedoppler} found out the physical relationship among target, first-order ghost and second-order ghost. The Hough transform was applied to find out real target in range-Doppler domain. Machine learning based methods have  also been adopted to recognize the ghost from measured observations. After training with the pre collected data from vehicle radar, the ghost can be distinguished successfully \cite{machinelearningghost}. Building a cooperative sensor network is one of the most effective method to suppress ghost. The authors in \cite{ghostfusion} analyzed the aspect properties of ghost target, and introduced a sensor network with multiple monostatic radars to mitigate the ghost and realize indoor tracking. In \cite{Junyang}, a distributed MIMO radar network was deployed, to sequentially separate the ghosts out.

The design of cooperative schemes, encompassing both the fusion method and node selection strategy, is also crucial for the detection and localization performance in a multiple-station network, especial for those resource limited scenarios. The geometric configuration of the participating station directly affects the detection and localization performance due to the difference in received signal intensity \cite{NETWORKLOCALIZATION}. An straightforward method is to reduce the resource cost on BS nodes, to increase the positioning accuracy while conserving power and reducing interference between sensing communication \cite{lth2024wcl}. Many investigations leverage the Cramer-Rao Lower Bound (CRLB), which provides a theoretical minimum for localization error, as a metric for optimizing the joint allocation of network resources like energy and spectrum  \cite{JASP} \cite{lth2024wcnc}. Nevertheless, the CRLB is not always achievable in low SNR conditions. To address this limitation, reinforcement learning (RL) based methods have been proposed. These approaches can optimize a non-closed-form localization metric, such as the mean square error (MSE) output by a specific algorithm, making them more practical for real-world scenarios with varying SNR \cite{RLallocation}, \cite{yang2021vtc}.

\subsection{Contribution}
To address aforementioned  challenges while fully leveraging current infrastructures, we propose a multi-layered, RF-based cooperative framework for UAV sensing, detection, and localization. Our approach begins with robust single-station sensing for LLS targets by combination methods of CA-CFAR detector and MDS based algorithm. Subsequently, measurements from multiple stations are intelligently fused using a grid-based probabilistic algorithm, which incorporates clustering to handle multi-UAV scenarios and effectively mitigate ghost targets.   To improve the network efficiency, we introduce a RL-based optimization strategy that adaptively balances the localization accuracy and the number of involved stations. The main contributions of this paper are summarized as follows.

\begin{itemize}
\item We first establish the framework for low altitude UAV sensing and detection, which is founded on a hybrid TOA and AOA localization method. The ghost types caused by indirect propagations are also analyzed comprehensively. 
\item A CA-CFAR algorithm is then proposed to suppress clutters and obtain the ranging and angle information from received signals. The MDS-based methodology using EMD is employed to recognize the LLS UAV and further improve detection performance.
\item Subsequently, a clustering based method is provided in multi-UAV scenarios, where a probabilistic grid-based data fusion method is proposed to mitigate the ghost targets and enhance the localization accuracy. 
\item The CRLB is derived analytically to illustrate the theoretical performance limits of this cooperative localization network. Two joint localization accuracy and node selection optimization problems are formulated and solved by RL-based algorithm to optimize the network efficiency. Numerical simulations and practical experiments are conducted to validate the superiority and feasibility of proposed framework.
\end{itemize}

The rest of this paper is organized as follows. Section II introduces the system model, detailing the measuring signal and ghost generation types. Section III presents algorithms of signal parameters estimation, MDS extraction and data fusion for UAV sensing, detection and localization, respectively. In section IV, the CRLB is derived, and two optimization problems are formulated and solved. Subsequently, simulation results and illustrated experiments are carried out in section V. Finally, Section VI draws the conclusion and key findings of this paper.

\section{System Model}
\subsection{Scenario}
\begin{figure}[t]
	\centering
	\includegraphics[width=1\columnwidth]{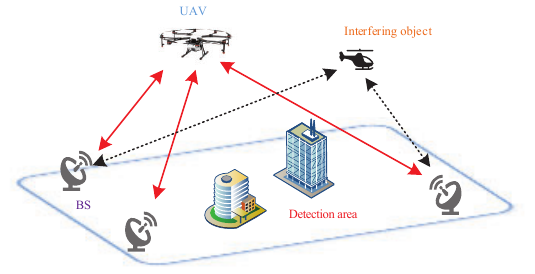}
	\caption{System model for multi-BS collaborative UAV detection in a complex low-altitude environment.}
	\label{fig:System_Model}
\end{figure}
As depicted in Fig.~\ref{fig:System_Model}, we consider a low-altitude surveillance scenario in the urban. There are $N_\text{BS}$ BSs deployed to monitor a specific area of interest. Within this area, there may be one or more non-cooperative UAVs to be detected and localized. There are also some other unintentional moving objects (e.g., helicopters, birds) and a number of randomly located interactive objects, such as high buildings and other cooperative aerial devices. These interactive objects act as reflectors and are the primary source of multipath propagation, which can generate ghost echoes. Each BS attempts to detect and localize UAVs based on received echoes. The primary focus of this work is to recognize and locate the UAV by intelligently fusing the observations from all individual stations. The investigation is carried out based on the following key assumptions:
\begin{itemize}
	\item The received echoes at BSs may be generated by the UAV, unintentional targets, and randomly located interactive objects. The presence and location of interactive objects may vary over time due to the dynamic nature of the environment.
	\item The interactive objects are treated as uncorrelated scatterers. For simplicity in modeling multipath, we primarily consider single-interaction bounces.
	\item During the detection interval (a single coherent processing interval), the UAVs are assumed to be quasi-static, meaning their velocity is constant and their position does not change significantly.
\end{itemize}

\subsection{Signal Model}
Generally, the received signal ${r(t)}$ at the BS, superposed by the signals from UAVs and ghosts, can be expressed as
\begin{align}
	\begin{split}
		r(t)=&\sum_{i=0}^{N_{\text{t}}}A_iS(t-\tau_i(t))\\
		&+\sum_{j=N_{\text{t}}+1}^{N_{\text{t}}+N_{\text{g}}+1}A_jS(t-\tau_j(t))+w(t),
		\label{eq:received_signal}
	\end{split}
\end{align}
where $S(t)$ represents the transmitted signal, and $w\left( t \right)$ is additive white Gaussian noise (AWGN). The first summation accounts for the echoes from the $N_{\text{t}}$ real UAVs, while the second part represents the echoes from the $N_{\text{g}}$ multipath ghosts in the environment. The term $A_i$ is the amplitude of the received signal from the $i$-th scatterer, which is determined by the radar range equation and expressed as
\begin{align}
	A_i=\sqrt{\frac{P_{\rm{tx}}G_{\rm{tx}}G_{\rm{rx}}\lambda^2\sigma_i}{(4\pi)^3R_i^4}}.
	\label{eq:amplitude}
\end{align}
Here, $P_{\rm{tx}}$ is the transmit power, $G_{\rm{tx}}$ and $G_{\rm{rx}}$ are the transmit and receive antenna gains, respectively, $\lambda$ is the signal wavelength, and $R$ is the distance to the scatterer. The term $\sigma_i$ indicates the RCS of the $i^\text{th}$ scatterer.

The term $\tau_i(t)$ is time delay of the signal reflected from UAV $i$, which is a function of its range and velocity. Assuming a UAV is at an initial distance $R_i$ from the BS and moving with a radial velocity $v_i$, the two-way time delay can be expressed as
\begin{equation}
	\tau_i \left( t \right) = 2\left( {R_i + v_it} \right)/c,
	\label{eq:time_delay}
\end{equation}
where $c$ is the speed of light in free space.

To enable angle estimation, we assume each BS is equipped with a uniform linear array (ULA) consisting of $K$ antenna elements. The signal received at the $k$-th element of the array, $r_k(t)$, can be modeled based on the signal at the first element and a phase shift corresponding to the angle of arrival
\begin{align}
	r_k(t)=r_0(t)\exp\big(j\frac{2\pi}{\lambda}d_{\rm{an}}(k-1)\sin\theta\big),
	\label{eq:ula_signal}
\end{align}
where $r_0(t)$ is the received signal at the first antenna element, and $d_{\rm{an}}$ is the distance between adjacent antenna elements, and $\theta$ is the AoA of the signal from the UAV.

\subsection{Ghost Target Generation}
In complex low-altitude environments, signals will propagate along multiple paths before returning to the receiver, leading to the generation of ``ghost'' targets, which are false detections.

\begin{figure}[htbp]
	\centering
	\includegraphics[width=0.9\columnwidth]{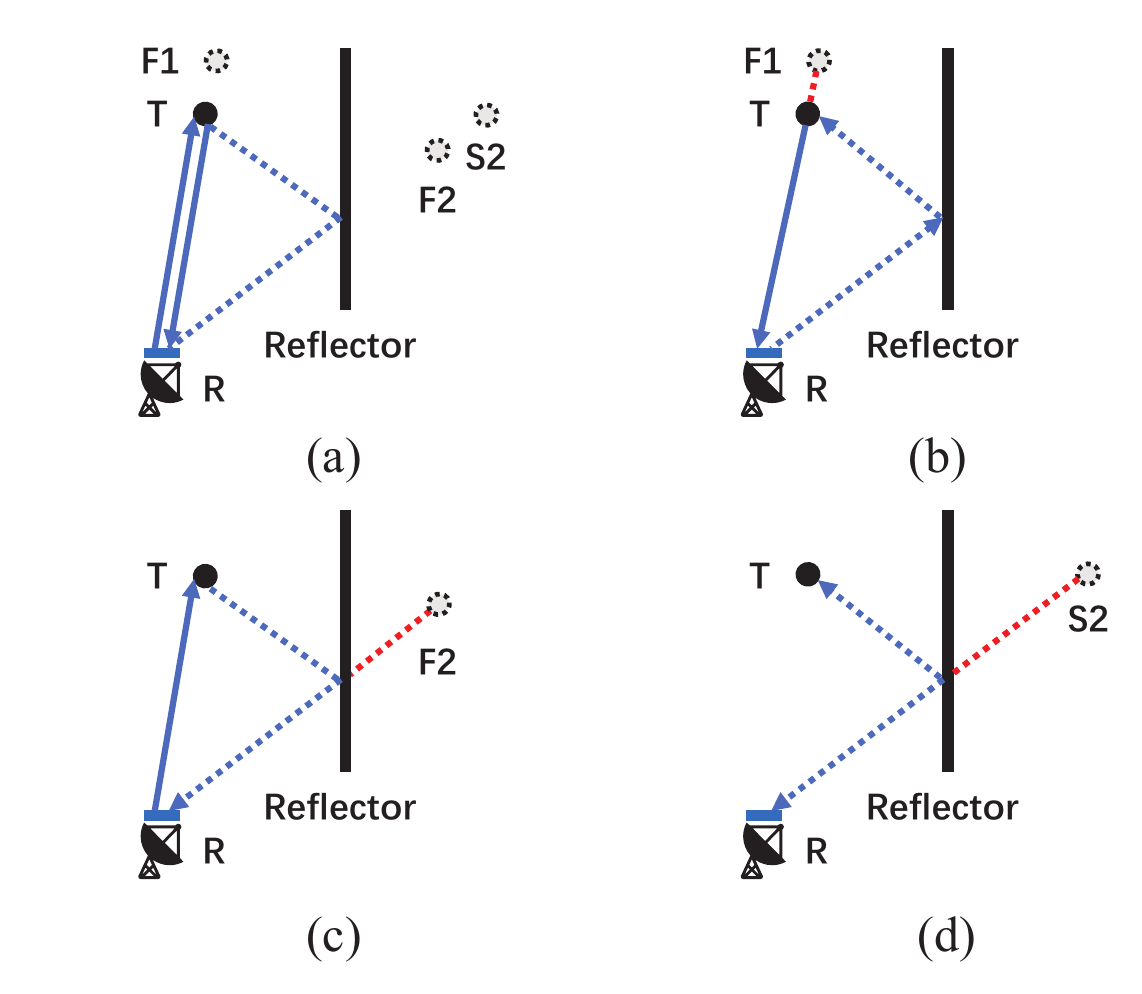}
	\caption{Geometry types of multipath ghost generation.}
	\label{fig:ghostgeomodel}
\end{figure}

To illustrate the properties of multipath ghosts, we consider a detection scenario with one BS, one UAV, and one reflector, as shown in Fig.~\ref{fig:ghostgeomodel}. The signal transmitted by the BS is assumed to be received by both the UAV and the reflector within its field of view (FoV). Ghosts are detected due to these indirect signal reflections. Based on the propagation path, these received signals can be categorized into the following types.

\begin{enumerate}
	\item \textbf{Direct Path:} As shown in Fig.~\ref{fig:ghostgeomodel}(a), the direct or line-of-sight path follows the propagation BS$\rightarrow$UAV$\rightarrow$BS. This corresponds to the real detection of the UAV.
	\item \textbf{Type-1 First-Order Ghost:} This ghost involves a reflection off the reflector before the signal illuminates the UAV. As shown in Fig.~\ref{fig:ghostgeomodel}(b), it experiences the propagation path BS$\rightarrow$Reflector$\rightarrow$UAV$\rightarrow$BS.
	\item \textbf{Type-2 First-Order Ghost:} This ghost, shown in Fig.~\ref{fig:ghostgeomodel}(c), occurs when the signal reflects off the reflector after scattering from the UAV. The propagation path is BS$\rightarrow$UAV$\rightarrow$Reflector$\rightarrow$BS.
	\item \textbf{Second-Order Ghost:} Fig.~\ref{fig:ghostgeomodel}(d) shows a second-order ghost, where both the transmitted and received signals bounce off the reflector. The path is BS$\rightarrow$Reflector$\rightarrow$UAV$\rightarrow$Reflector$\rightarrow$BS\footnote{If multiple reflectors exist, a signal may experience more than two indirect reflections. Generally, the energy of these higher-order ghosts is relatively small, making them less likely to be detected.}.
\end{enumerate}

The detection probability of both the real UAV and the multipath ghosts is highly related to their signal-to-noise ratio (SNR) at the receiver. For the signal reflected from the direct path, the SNR is expressed as
\begin{align}
	{\gamma_t}=\frac{P_{\rm{t}}}{N_0}=\frac{P_{\rm{tx}} G_{\rm{tx}}}{4 \pi  R^{2}} \cdot \sigma_{\rm{t}} \cdot \frac{1}{4 \pi R^{2}} \cdot \frac{G_{\rm{rx}} \lambda^{2}}{4 \pi} \cdot \frac{1}{N_0},
\end{align}
where ${P_{\rm{t}}}$ is the input power of signal from target UAV, $N_0$ is the noise power, $P_{\rm{tx}}$ is the output power from the transmitter, $R$ is the distance from the BS to the UAV, $\lambda$ is the BS wavelength, $\sigma_{t}$ is the non-fluctuating RCS of the UAV, and $G_{\rm{tx}}$ and $G_{\rm{rx}}$ are the antenna gains of the transmitter and receiver, respectively.

Conversely, the power of a signal received from an indirect path is lower than direct path due to the additional path loss and reflection losses. We typically only consider first-order ghosts, as higher-order ghosts have significantly lower energy. Assuming the distances from the reflector to the UAV and the BS are $R_{1}$ and $R_{2}$ respectively, and the RCS of the reflector is $\sigma_{r}$. Therefore, the SNR for an indirect path ghost can be expressed as
\begin{align}
	{\gamma_g}=\frac{P_{\rm{tx}} G_{\rm{tx}}}{4 \pi  R^{2}} \cdot \sigma_{t} \sigma_{r} \cdot \frac{1}{{4 \pi}^{2} {{R_{1}}^{2}}{{R_{2}}^{2}}} \cdot \frac{G_{\rm{rx}} \lambda^{2}}{4 \pi} \cdot \frac{1}{N_0}.
\end{align}

\section{UAV Detection and Localization Algorithm}
\begin{figure*}[ht]
	\centering
\vspace{-5mm}
	\includegraphics[width=1\textwidth]{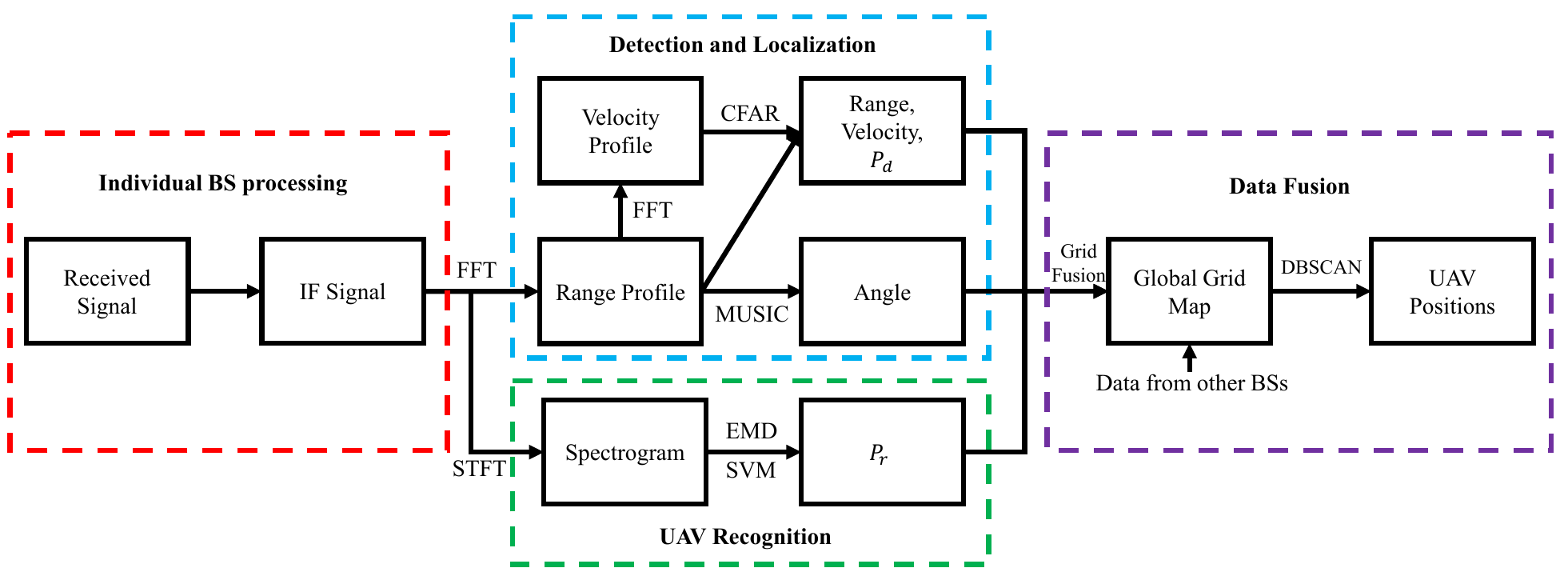} 
	\caption{The end-to-end framework for multi-BS collaborative UAV detection and localization.}
\vspace{-5mm}
	\label{fig:overall_framework}
\end{figure*}

This section introduces the comprehensive process of our proposed UAV Detection and Localization framework, where the parameter estimation, MDS extraction, multi-station data fusion, are elaborated, respectively. The end-to-end workflow is illustrated in Fig.~\ref{fig:overall_framework}. 
\subsection{Range, velocity and angle estimation}
In this paper, we utilize the Linear Frequency Modulated Continuous Wave (LFMCW) as the transmitted signal $S(t)$, whose generation is compatible with the hardware capabilities of future BS, and is expressed as
\begin{equation}
	{S}\left( t \right) = {A}\exp \left\{ {j2\pi \left( {{f_0}t + \mu {t^2}/2} \right)} \right\},
	\label{eq_transmitted_signal}
\end{equation}
where ${f_0}$ is the carrier frequency, $\mu  = B/T$ is the linear modulation slope for a signal with bandwidth $B$ and duration $T$, and ${A}$ is the amplitude. At each BS, the range, angle, and velocity of UAV are obtained from the received echoes through following signal processing steps.
\subsubsection{Intermediate Frequency (IF) Signal Generation}
The IF signal is achieved by mixing the received echoes in (\ref{eq:received_signal}) with a local copy of the transmitted signal. This process, known as de-chirping, shifts the echoes from different scatterers to baseband, where each scatterer's echo becomes a sinusoidal tone. The resulting baseband IF signal ${r_{\rm{IF}}}\left( t \right)$ is written as
\begin{align}
	\begin{split}
		r_{\rm{IF}}(t) = &\sum_{i=0}^{N_{\text{t}}}A_iS_{{\rm IF}_i}(t-\tau_i(t))\\
		&+\sum_{j=N_{\text{t}}+1}^{N_{\text{t}}+N_{\text{g}}+1}A_jS_{\rm{IF_j}}(t-\tau_j(t)) + \tilde{w}(t),
	\end{split}
\end{align}
where ${S_{{\rm{IF}}_i}}\left( t \right)$ represents the IF component corresponding to the $i$-th scatterer, and $\tilde w\left( t \right)$ is the mixer noise. The frequency of each IF component ${S_{{\rm{IF}}_i}}\left( t \right)$ is proportional to the round-trip delay $\tau_i(t)$, which in turn encodes its range and velocity.

\subsubsection{IF Signal Digitalization}
The analog IF signal is then sampled to produce a discrete-time signal. Denoting sampling period as ${T_s} = T/L$, the sampling time is $t = mT + lT/L$, where $l \in \left\{ {0,...,L - 1} \right\}$ and $m \in \left\{ {0,...,M - 1} \right\}$ represent the fast-time (for range) and slow-time (for velocity) domains, respectively. The discrete received signal is expressed as
\begin{equation}
	{s_{\rm{r}}}\left( {l,m} \right) = {r_{\rm{IF}}}\left( {mT + lT/L} \right){\rm{ + }}\tilde w\left( {mT + lT/L} \right).
\end{equation}
This forms a matrix ${{\bf{S}}_{\rm{r}}} \in {\Re ^{L \times M}}$.

\subsubsection{Range Profile Generation}
A range profile is generated by performing a Fast Fourier Transform (FFT) along the fast-time domain. A Hamming window is typically applied before the FFT to reduce sidelobes, which means the rang matrix 
\begin{equation}
	{\bf{R}} = \left( {{{\bf{W}}_L}{{\bf{F}}_L}} \right) \cdot {{\bf{S}}_{\rm{r}}} \in {\Re ^{L \times M}},
\end{equation}
where ${{\bf{W}}_L}$ is the windowing matrix and ${{\bf{F}}_L}$ is the FFT matrix. Therefore, the range corresponding to the $l$-th bin is estimated as
\begin{equation}
	R\left( l \right) = \frac{{c{f_s}l}}{{2L\mu }},
\end{equation}
where ${f_s} = 1/{T_s}$ is the sampling frequency.

\subsubsection{Velocity Profile Generation}
Similarly, a velocity profile is obtained by performing an FFT across the slow-time domain, and is expressed as
\begin{equation}
	{\bf{V}} = {\bf{R}} \cdot \left( {{{\bf{W}}_M}{{\bf{F}}_M}} \right) \in {\Re ^{L \times M}}.
\end{equation}
The velocity corresponding to the $m$-th bin is estimated as
\begin{equation}
	V\left( m \right) = \frac{{cm}}{{2{f_0}MT}}.
\end{equation}

\subsubsection{Range and Velocity Determination}
Subsequently, a CA-CFAR detection strategy is adopted to distinguish UAV echoes from background noise in the range-velocity map \cite{Richards2014Fundamentals}. The main steps are listed in Algorithm~\ref{alg:CFAR_estimation}. A closed-form of the detection probability, $P_{\rm{d},\rm{t}}$ for a specific UAV/ghost, can be achieved by calculating the probability that the signal envelope is over the detection threshold \cite{richards2005fundamentals} as
\begin{align}
	{P_{\rm{d},\rm{t}}} = {\left( {1 + \frac{{\left( {{N_{\rm{R}}} {N_{\rm{V}}}} \right)\left( {P_{{\rm{FA}}}^{ - {1 \mathord{\left/
									{\vphantom {1 {\left( {{N_{\rm{R}}} \cdot {N_{\rm{V}}}} \right)}}} \right.
									\kern-\nulldelimiterspace} {\left( {{N_{\rm{R}}} {N_{\rm{V}}}} \right)}}} - 1} \right)}}{{\left( {{N_{\rm{R}}} {N_{\rm{V}}}} \right)\left( {1 + \gamma_{\rm{t}}} \right)}}} \right)^{ - {N_{\rm{R}}} {N_{\rm{V}}}}},
\end{align}
where $P_{\text{FA}}$ is the desired false alarm rate, and $N_{\text{R}}$ and $N_{\text{V}}$ are the number of reference cells in range and velocity.

\begin{algorithm}[H]
	\caption{The CFAR-based range and velocity estimation}
	\label{alg:CFAR_estimation}
	\begin{algorithmic}[1]
		\STATE \textbf{Initialization:} Determine the false alarm probability ${P_{\rm{FA}}}$, the number of reference units ${N_{\rm{R}}}$ in range and ${N_{\rm{V}}}$ in velocity, set estimation index $q=0$.
		\FOR{$l = 0:L-1,m = 0:M-1$}
		\STATE {Compute the mean signal power $\beta _{l,m}$ within the reference windows.}
		\STATE {Determine the threshold $\gamma_{\rm{th}} $ using
			$\gamma_{\rm{th}}  = \left( {{N_{\rm{R}}} \cdot {N_{\rm{V}}}} \right)\left( {P_{{\rm{FA}}}^{ - {1 \mathord{\left/
							{\vphantom {1 {\left( {{N_{\rm{R}}} \cdot {N_{\rm{V}}}} \right)}}} \right.
							\kern-\nulldelimiterspace} {\left( {{N_{\rm{R}}} \cdot {N_{\rm{V}}}} \right)}}} - 1} \right)\beta _{l,m}$.
		}
		\IF{the power of detection unit ${E_{l,m}} \ge \gamma_{\rm{th}} $}
		\STATE {$\hat{{R}}\left( q \right) = R\left( l \right),\hat{{V}}\left( q \right) = V\left( m \right)$}.
		\STATE {$q = q + 1$}.
		\ENDIF
		\ENDFOR
		\STATE \textbf{Output:}\ $\hat{{{\bf{R}}}}, \hat{{\bf{V}}}$.
	\end{algorithmic}
\end{algorithm}

\subsubsection{Angle Estimation}
As established in the signal model, each BS is equipped with a ULA, allowing it to estimate the AoA of the signals reflected from a UAV. Off-the-shelf estimator as multiple signal classification (MUSIC) \cite{Music2016algorithm} can be used. The angle estimates for these two algorithms are expressed as 
\begin{equation}
	{\hat \theta _{{\rm{MUSIC}}}}{\rm{ = }}\mathop {\arg \min }\limits_\theta  {\kern 1pt} {\bf{a}}{\left( \theta  \right)^{\rm{H}}}{{\bf{U}}_{\rm{N}}}{\bf{U}}_{\rm{N}}^{\rm{H}}{\bf{a}}\left( \theta  \right),
\end{equation}
where ${{\bf{U}}_{\rm{N}}}$ is the noise subspace matrix and ${\bf{a}}\left( \theta  \right)$ is the array steering vector.
These estimated parameters including range, velocity, angle, and detection probability are then available for multi-BS fusion and further analysis.

\subsection{Micro-Doppler Feature Enhancement for UAV Recognition}
While the CA-CFAR detector can identify the presence of moving objects, it cannot distinguish a UAV from other objects. To achieve robust UAV recognition, we leverage the unique MDS generated by the rotating blades of a UAV. These micro-motions produce a distinct frequency modulation on the reflected echo signal, which can be used as a feature for classification \cite{chen2006micro}.
\begin{figure}[t]
	\centering
	\includegraphics[width=0.5\columnwidth]{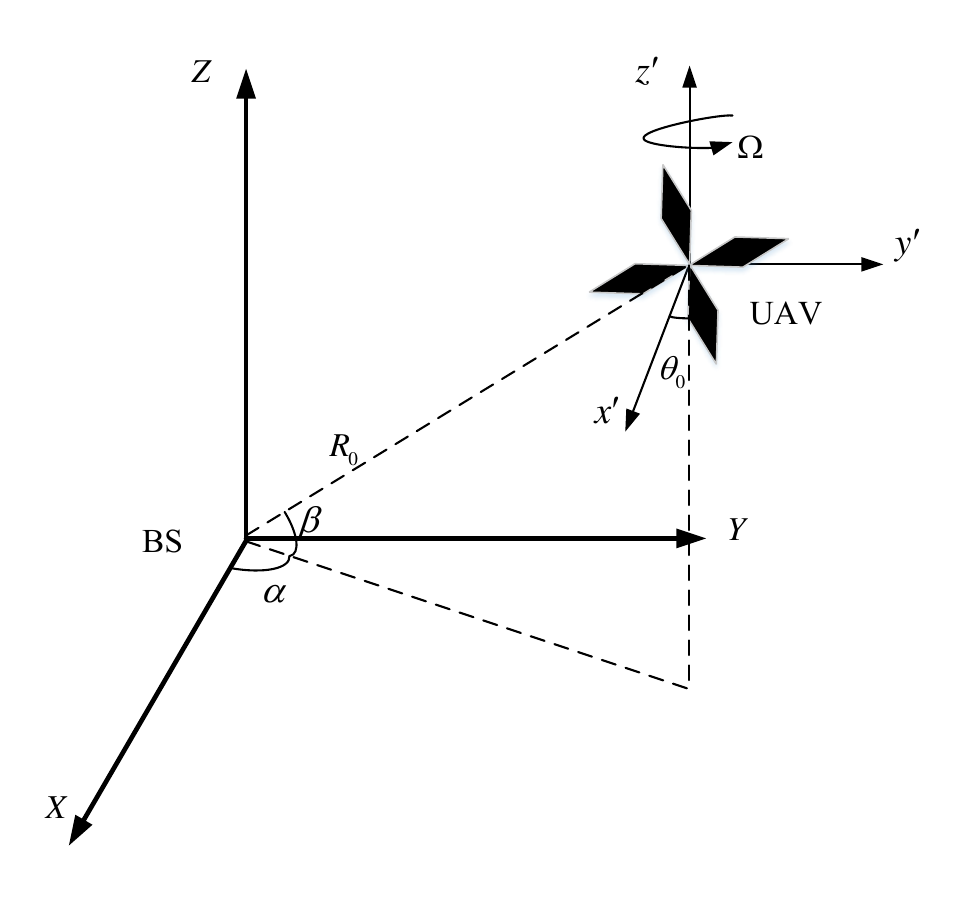}
	\caption{Geometric model for analyzing the micro-Doppler effect from a UAV's rotor blade as observed by a BS.}
	\label{fig:Micro-doppler_Model}
\end{figure}
To model this effect, we consider a UAV with ${N_{{\rm{rot}}}}$ rotors. The echoes from a single rotor can be modeled as the coherent sum of reflections from its ${N_{{\rm{bl}}}}$ blades \cite{Micro2018,Chen2011The}. The echo from one rotor ${{s_{{n_{{\rm{rot}}}}}}\left( t \right)}$ can be expressed as
\begin{equation}
	\begin{array}{l}
		{{s_{{n_{{\rm{rot}}}}}}\left( t \right)} = {L_{{\rm{bl}}}}e^{  { - j\frac{{4\pi }}{\lambda }{R_0}} }{\kern 1pt}  \\\sum\limits_{{n_{{\rm{bl}}}} = 0}^{{N_{{\rm{bl}}}} - 1}{\rm{sinc}} {\left\{ {\frac{{2\pi {L_{{\rm{bl}}}}}}{\lambda }\cos \beta \cos \left( {{\theta _{{n_{{\rm{bl}}}}}} + \Omega t - \alpha } \right)} \right\}} e^ { { - j{\Phi _{{n_{{\rm{bl}}}}}}\left( t \right)} },
	\end{array}
\end{equation}
where ${L_{{\rm{bl}}}}$ is the length of a single blade. As shown in Fig.~\ref{fig:Micro-doppler_Model}, ${R_0}$ is the distance from the rotor's center of rotation to the BS, $\Omega$ is the rotational angular frequency of the blade, and $\alpha$ and $\beta$ are the azimuth and elevation angles of the rotor relative to the BS line-of-sight (LoS), respectively. The term ${\Phi _{{n_{{\rm{bl}}}}}}\left( t \right)$ represents the phase function of the $n_{\text{bl}}$-th blade.
The total echo signal from the UAV ${S_\Sigma }\left( t \right)$ is the sum of echoes from all its rotors, 
\begin{equation}
	{S_\Sigma }\left( t \right) = \sum\limits_{{n_{{\rm{rot}}}} = 1}^{{N_{{\rm{rot}}}}} {{s_{{n_{{\rm{rot}}}}}}\left( t \right)}.
\end{equation}

To classify the UAV based on this signal, we employ a process involving EMD and a SVM classifier. The full procedure for acquiring the recognition probability ${P_{\rm{r}}}$, is described in Algorithm~\ref{alg:Recognition_probability}. A STFT is first applied to ${S_\Sigma }\left( t \right)$ to generate a time-frequency spectrogram to represent the MDS. The EMD algorithm then decomposes the signal into a set of intrinsic mode functions (IMFs). Features such as zero-crossing rate, normalized energy, and standard deviation are extracted from the first four IMFs. These features form a vector that is fed into a pre-trained SVM classifier, which outputs the recognition probability ${P_{\rm{r}}}$. This probability quantifies the confidence that the detected object is indeed a UAV.
\begin{algorithm}[h]
	\caption{Recognition Probability Acquisition}
	\label{alg:Recognition_probability}
	\begin{algorithmic}[1]
		\STATE \textbf{Input:} Total echo signal from $t$-th UAV, ${S_\Sigma }$.
		\STATE \textbf{Step 1:} Convert the echo signal to a real signal ${x_\Sigma } = \left| {{S_\Sigma }} \right| \in \Re $.
		\STATE \textbf{Step 2:} Decompose ${x_\Sigma }$ into ${x_\Sigma } = \sum\limits_{d = 1}^D {{m_d}}  + {q_D}$ using the EMD algorithm, where ${m_d}$ is the $d$-th Intrinsic Mode Function (IMF) and ${q_D}$ is the residual.
		\STATE \textbf{Step 3:} Extract features (zero-crossing number, normalized signal energy, standard deviation, and zero-crossing ratio) from the first four IMFs.
		\STATE \textbf{Step 4:} Obtain the recognition probability ${P_{\rm{r},\rm{t}}}$ by feeding the feature vector into an SVM classifier with a radial basis function (RBF) kernel.
		\STATE \textbf{Output:} Recognition probability, ${P_{\rm{r},\rm{t}}}$.
	\end{algorithmic}
\end{algorithm}
\subsection{Multi-Station Data Fusion for Localization}
While a single BS can estimate the range, velocity, and angle of a UAV, its observations are susceptible to errors from noise, clutter, and multipath ghosts. To overcome these limitations and achieve robust, high-accuracy localization, we fuse observations from multiple spatially BSs. A grid-based probabilistic data fusion framework is proposed to reduce interference and enhances the reliability of the final localization. The process involves grid initialization and calibration, data fusion, and clustering.

\subsubsection{Grid Initialization and Calibration}
The first step is to discretize the surveillance area into a grid and to map the local observations from each BS into this common, unified world coordinate system. This ensures that all measurements can be compared and combined meaningfully.

\begin{figure}[t]
	\centering
	\includegraphics[width=0.8\columnwidth]{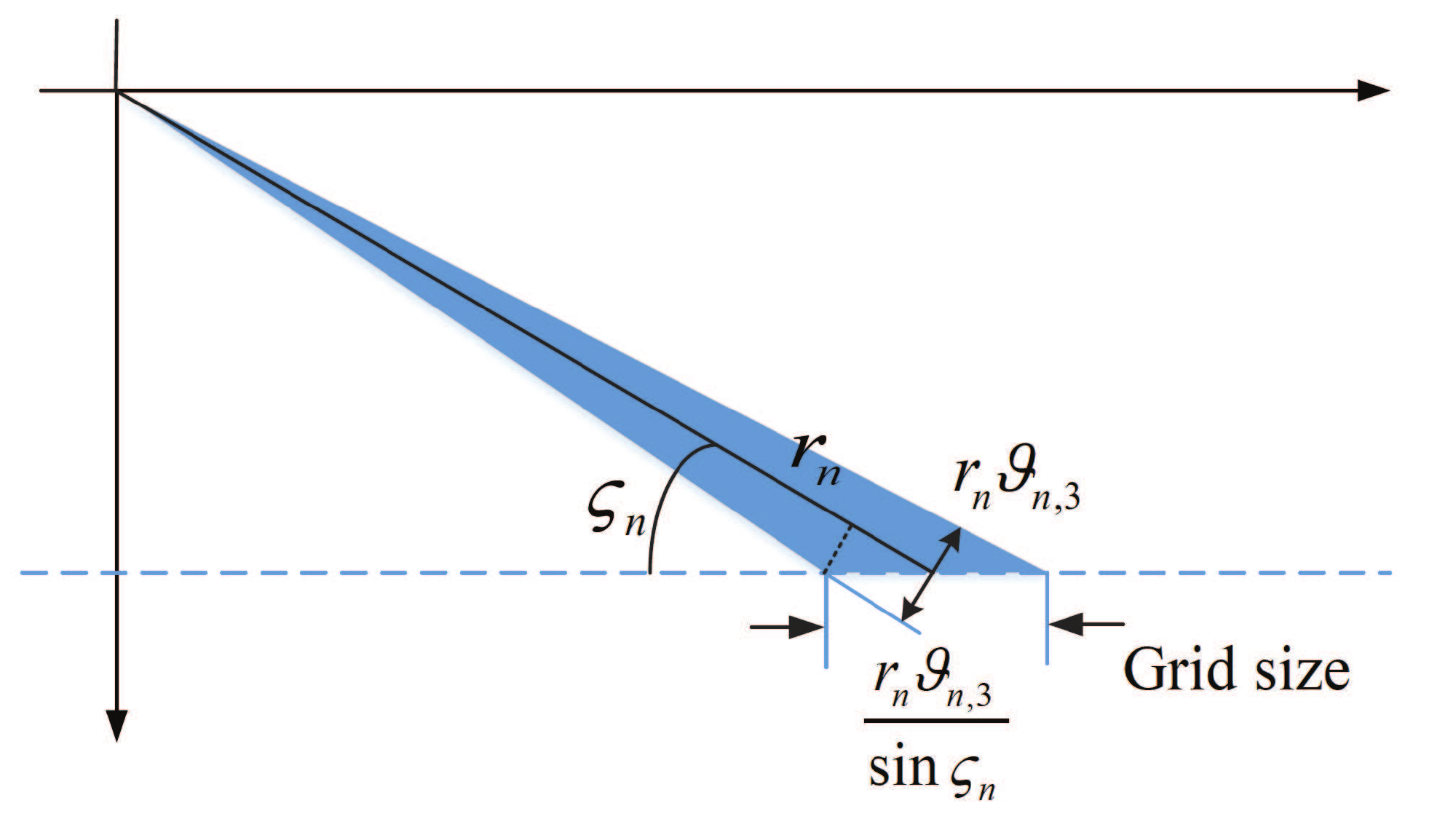}
	\caption{Determination of the possible detection region.}
	\label{fig:Grid_Size}
\end{figure}

A critical aspect of this process is determining the size of the grid cells, which should be related to the resolution of the sensing BSs. For a given BS, a measurement corresponds not to an infinitesimal point but to a ``possible detection region'', the size of which is determined by its range and angular resolution. As shown in Fig.~\ref{fig:Grid_Size}, the size of this region $h_n$ for the $n$-th BS is the minimum of its effective range and beam resolutions with
\begin{align}
	h_{n}=\min\{h_{n,\text{range}}, h_{n,\text{beam}}\},
\end{align}
where $h_{n,\text{range}}$ and $h_{n,\text{beam}}$ are defined as:
\begin{align}
	h_{n,\text{range}}&=\underbrace{\Delta R}_{\text{range resolution}}/\cos{\varsigma_n}=\frac{c}{2B}/\cos{\varsigma_n},\\
	h_{n,\text{beam}}&=r_n\underbrace{\vartheta_{n,3}}_{\text{angle resolution}}/\sin {\varsigma_n}.
\end{align}
Here, $\Delta R$ is the intrinsic range resolution (dependent on bandwidth), and $\vartheta_{n,3}$ is the 3dB beamwidth of the antenna, and $r_n$ is the measured range, and $\varsigma_n$ is the transmission angle. The final grid size for the entire map is chosen to be the minimum resolution across all participating BSs to avoid loss of information.

Once the grid is established, each local measurement is calibrated. Let the position of the $n$-th BS in the world be $\mathbf{x}_{n}$=$(x_{n},y_{n})$, with a rotation angle $\psi_n$. If this BS detects a UAV at range $r_n$ and angle $\theta_n$, the UAV's world position, $\mathbf{x}_{tn}$, is calculated as
\begin{align}
	\left[
	\begin{array}{c}
		x_{tn}\\
		y_{tn}
	\end{array}
	\right]
	=
	\left[
	\begin{array}{c}
		x_{n}\\
		y_{n}
	\end{array}
	\right]
	+r_n
	\left[
	\begin{array}{c}
		\cos(\psi_n+\theta_n)\\
		\sin(\psi_n+\theta_n)
	\end{array}
	\right].
\end{align}
This mapping allows us to translate every detection from its local polar coordinates to the corresponding grid cells in the global map.

\subsubsection{Data Fusion on a Grid Map}
After calibration, the observations are fused on the grid map. This probabilistic approach combines information while filtering out inconsistent measurements, such as those from ghosts.

The event that grid cell $i$ is occupied by a UAV is defined as $O_i$. When the $n$-th BS receives an echo that maps to grid cell $i$ (event $y_i^{(n)}$), the posterior probability of the cell being occupied is updated using the Bayesian theorem as
\begin{align}
	p(O_i|y_i^{(n)})=p(O_i)\cdot p(y_i^{(n)}|O_i).
\end{align}
The likelihood term $p(y_i^{(n)}|O_i)$ is the recognition probability $P_{\rm{r}}$ from the micro-Doppler analysis. This assigns higher likelihood to detections that exhibit UAV-like characteristics, effectively suppressing echoes from unintentional objects.

To combine observations from all $N_{\text{BS}}$ BSs, we adopt a log-likelihood ratio probability fusion framework. For the $n$-th BS, the occurrence probability of grid cell $i$ is updated iteratively
\begin{align}
	p(O_i)^{(n)}=\text{log}\big(\frac{p(O_i|y_i^{(n)})}{P_{\text{FA}}}\big)+p(O_i)^{(n-1)}.
\end{align}
After iterating through all BSs, the final fused grid map contains enhanced probabilities for true UAV locations. The complete process is detailed in Algorithm~\ref{alg:multi_radar_fusion}. The final UAV position, $\hat{\mathbf{p}}$, can then be estimated using a minimum mean square error (MMSE) as
\begin{align}
	\hat{\mathbf{p}}=\sum_{i=1}^{N_\text{t}}p(O_i)^{(N_{\text{BS}})}l_{i},
\end{align}
where $l_i$ is the location of the $i$-th grid cell.

\begin{algorithm}[ht]
	\caption{Grid Map Based Multi-BS Fusion}
	\label{alg:multi_radar_fusion}
	\begin{algorithmic}[1]
		\STATE {\bf Initialization:} Number of BSs $N_{\text{BS}}$, grid point locations $l_i$, BS locations and orientations $\mathbf{x}_{n}, \psi_n$, and initial probabilities $p(O_i)$.
		\FOR {BS $n=1:N_{\text{BS}}$}
		\STATE{For each local measurement ($r_n, \theta_n$), determine the possible detection region $\mathbf{R}$ on the grid map.}
		\FOR {each grid point $i\in \mathbf{R} $}
		\STATE {$p(O_i)^{(n)}=\text{log}\big(\frac{p(O_i|y_i^{(n)})}{P_{\text{FA}}}\big)+p(O_i)^{(n-1)}$}
		\ENDFOR
		\ENDFOR
		\STATE {\bf Normalization:} Normalize the final probabilities across the grid.
		\STATE {\bf Output:} MMSE-based position estimate $\hat{\mathbf{p}}$.
	\end{algorithmic}
\end{algorithm}

\subsubsection{Clustering for Multiple UAVs}
In scenarios with multiple UAVs, the fused grid map may contain several high-probability regions. To distinguish individual UAVs, we use the Density-Based Spatial Clustering of Applications with Noise (DBSCAN) method. DBSCAN groups high-probability grid cells into distinct clusters, each corresponding to a different UAV while identifying sparse noise points \cite{lthiotj}. The algorithm is detailed in Algorithm~\ref{alg:DBSCAN_clustering}.

\begin{algorithm}[ht]
	\caption{Clustering using DBSCAN}
	\label{alg:DBSCAN_clustering}
	\begin{algorithmic}[1]
		\STATE \textbf{Initialization:} The fused localization map ${\bf{M}}$, the radius of the Epsilon area (Eps), and the minimum number of neighbors (${N_{{\rm{Pts}}}}$).
		\FOR{each point ${\bf{P}}$ in ${\bf{M}}$ not yet visited}
		\STATE{Mark ${\bf{P}}$ as visited.}
		\STATE{Find all neighbor points of ${\bf{P}}$ within Eps.}
		\IF{the number of neighbors is less than ${N_{{\rm{Pts}}}}$}
		\STATE{Mark ${\bf{P}}$ as noise.}
		\ELSE
		\STATE{Create a new cluster ${C}$ and add ${\bf{P}}$ to it.}
		\STATE{For each neighbor point ${\bf{P'}}$ of ${\bf{P}}$:}
		\STATE{\quad If ${\bf{P'}}$ is not visited, mark it as visited and find its neighbors. If it has enough neighbors, add them to a queue to be processed.}
		\STATE{\quad If ${\bf{P'}}$ is not yet a member of any cluster, add ${\bf{P'}}$ to cluster ${C}$.}
		\ENDIF
		\ENDFOR
	\end{algorithmic}
\end{algorithm}

\section{Performance Analysis and Optimization}

\subsection{The Cramer-Rao Lower Bound}
To evaluate the fundamental performance limits of the proposed multi-BS localization system, we utilize the CRLB, which provides a theoretical lower bound on the MSE for any unbiased estimator of a UAV's position. According to the Fisher information inequality, the MSE of the position estimate, $\hat{\mathbf{p}}$, is bounded as follows
\begin{align} \mathbb{E}\{||\hat{\mathbf{p}}-\mathbf{p}||^2\}\geq\mathcal{P}(\mathbf{p})=\text{tr}\{\mathbf{J}_{\rm{e}}(\mathbf{p})^{-1}\},
\end{align}
where $\mathbf{J}_{\rm{e}}(\mathbf{p})$ is the equivalent Fisher information matrix (EFIM) for the entire system \cite{shen2010fundamental}. The EFIM is constructed by aggregating the information from each antenna element of every participating BS. We can state the following proposition regarding its structure.

\textit{Proposition 1:} If there are $N_\text{BS}$ BSs deployed, and each BS is equipped with a $K$-element uniform linear antenna array, the equivalent EFIM of the UAV's position can be expressed as the sum of the individual information matrices from each antenna's ToA and AoA measurements:
\begin{align}
\mathbf{J}_{\rm{e}}(\mathbf{p})&=\sum_{n=1}^{N_{\text{BS}}}\sum_{k=1}^{K}\big(\mathbf{J}_{nk}^\text{TOA}(\mathbf{p})+\mathbf{J}_{nk}^\text{AOA}(\mathbf{p})\big),
\end{align}
where $\mathbf{J}_{nk}^\text{TOA}(\mathbf{p})$ and $\mathbf{J}_{nk}^\text{AOA}(\mathbf{p})$ are the FIM for the ToA and DoA measurements, respectively, from the $k$-th antenna element of the $n$-th BS.

\textit{Proof:} The detailed derivation of the Fisher Information Matrix is provided in the Appendix.

This proposition highlights two key remarks.
\begin{itemize}
	\item At each individual BS, the measurement information obtained from ToA ($\mathbf{J}^\text{TOA}(\mathbf{p})$) and DoA ($\mathbf{J}^\text{AOA}(\mathbf{p})$) can be linearly accumulated. The quality of this information is determined by system parameters such as transmit power and signal bandwidth, as well as the geometric relationship between the BS and the UAV.
	\item The measurement information from multiple BSs can also be linearly accumulated. This demonstrates that observations from all $N_\text{BS}$ BSs can be combined to achieve a more accurate and reliable position estimate compared to what is possible with a single BS.
\end{itemize}

The CRLB serves as an essential performance benchmark, allowing us to quantify how close the accuracy of our proposed fusion and localization algorithms comes to the theoretical best-case scenario.

\subsection{Reinforcement Learning for Station Selection}
In a dense network of BSs, it is often inefficient and unnecessary to activate all of them for UAV detection. Activating only the most useful BSs can significantly reduce energy consumption and computational load while maintaining or even improving localization accuracy. This section formulates the BS selection task as an optimization problem and presents a RL-based solution. We first consider a single UAV scenario and then extend the framework to handle multiple UAVs.

\subsubsection{Single UAV Scene}
\paragraph{Problem Formulation}
For a single UAV scene with a total of $N_\text{BS}$ available BSs, our objective is to select a subset of these BSs to find a balance between minimizing the localization MSE and the resource cost (number of active BSs). With the true position of the UAV being $\mathbf{p}_t$ and the estimated position being $\hat{\mathbf{p}_t}$, the MSE is $\mathcal{M}(\mathbf{R},\mathbf{p}_t)$. The optimization problem can then be written as
\begin{align}
	\text{$\mathscr{P}_1$: } \min &\quad \tau_{e}\mathcal{M}(\mathbf{R},\mathbf{p}_t)+\tau_{r} R_s  \\
	\nonumber
	\text{s.t.} &\quad \mathbf{R}=\left\{ r_1,r_2,...,r_{N_\text{BS}}\right\},\\
	\nonumber
	&\quad r_{ n} \in\{0,1\},
\end{align}
where $\mathbf{R}$ is the selection vector, with $r_n=1$ indicating that the $n$-th BS is selected. $R_s$ is the total number of selected BSs. $\tau_e$ and $\tau_r$ are the weighting factors for the MSE and resource cost, respectively. Since the MSE cannot be easily formulated as a closed-form function of the BS selection state $\mathbf{R}$, we adopt a non-parametric, model-free learning strategy.

\paragraph{Node Selection Strategy}
We employ Q-learning to find an optimal policy for BS selection. The Q-learning framework is defined within a Markov Decision Process (MDP) as follows:
\begin{itemize}
	\item \textbf{Agent}: The agent is the BS sensor network responsible for positioning.
	\item \textbf{State}: The state is the power allocation status for all BSs, represented by a binary vector $\mathbf{R}$ indicating which BSs are currently active (on/off).
	\item \textbf{Action}: An action is the decision to change the state, i.e., to select a new subset of BSs to activate for the next measurement.
	\item \textbf{Environment}: The environment is the physical space containing the UAV. The agent's interaction with the environment yields an observation, which is the localization result obtained using the fusion framework from the previously selected BSs.
\end{itemize}

The Q-learning algorithm iteratively updates a Q-table, which estimates the long-term reward of taking a certain action from a given state. The agent explores different selection strategies using an $\epsilon$-greedy policy. The reward function is designed to reflect the objective function $\mathscr{P}_1$
\begin{align}
	r=C_{0}-\tau^{e}\mathcal{M}-\tau^{r} R_s,
\end{align}
where ${C_{0}}$ is a constant baseline reward. By maximizing this reward, the agent learns to select BS configurations that result in a good trade-off between positioning error and resource cost. The complete training and regression process is outlined in Algorithm~\ref{alg:QL_selection}. To improve real-time performance, a pre-trained Back Propagation (BP) neural network can be used to approximate the Q-function, enabling rapid selection of the sub-optimal action based on a coarse initial UAV position estimate.

\begin{algorithm}[h]
	\caption{Reinforcement Learning Based BS Selection}
	\label{alg:QL_selection}
	\begin{algorithmic}[1]
		\STATE \textbf{Initialization:}
		\STATE \quad{A sensor network with $N_\text{BS}$ BSs with locations $\mathbf{p}_1,\mathbf{p}_2,...\mathbf{p}_{N_\text{BS}}$ and one single UAV with location $\mathbf{p}_t$. The estimated position of the UAV is $\hat{\mathbf{p}_t}$ according to Algorithm~\ref{alg:multi_radar_fusion}.}
		\STATE \quad{Generate the training dataset with $N_{train}$ UAV positions, expressed as $N_{train}={1,2,..,N_{train}}$.}
		\STATE \quad Define the RL values: i) State $S$, a binary series representing the on-off of BS nodes. ii) Reward $r$, affected by the localization error and number of enabled nodes. iii) Action $A$, the selection or abandonment of BS nodes.
		\STATE \textbf{Training:}
		\REPEAT
		\STATE {Initialize the Q-table.}
		\FOR {each episode of Q-learning}
		\IF {current iteration $\leq$ upper limit}
		\STATE {Select a set of actions with an $\varepsilon$-greedy policy.}
		\STATE {Update the Q-table and current state.}
		\ENDIF
		\ENDFOR
		\UNTIL {all $N_{train}$ data are processed}
		\STATE \textbf{Regression:}
		\STATE \quad{Input the coarse estimated UAV position $\hat{\mathbf{p}_{tc}}$.}
		\STATE \quad{Derive the sub-optimal action with the pre-trained BP network: $\hat{A}=$ BP network $(\hat{p_{tc}})$.}
	\end{algorithmic}
\end{algorithm}

\subsubsection{Multi-UAV Scene}
\paragraph{General Extension}
When multiple UAVs are present, the objective function is modified to account for the localization error of all $M$ UAVs. The optimization problem, $\mathscr{P}_2$, becomes minimizing the sum of weighted MSEs and the resource cost
\begin{align}
	\text{$\mathscr{P}_2$: } \text{min} &\quad \sum_{m=1}^{M} \tau_{e}\mathcal{M}_m(\mathbf{R},\mathbf{p}_t)+\tau_{r} R_s  \\
	\nonumber
	\text{s.t.} &\quad \mathbf{R}=\left\{ r_1,r_2,...,r_{N_\text{BS}}\right\},\\
	\nonumber
	&\quad r_{ n} \in\{0,1\}.
\end{align}
To satisfy this requirement, the reward function in the Q-learning algorithm must be adjusted to reflect this new objective.

\begin{figure}[htbp]
	\centering
	\includegraphics[width=0.8\columnwidth]{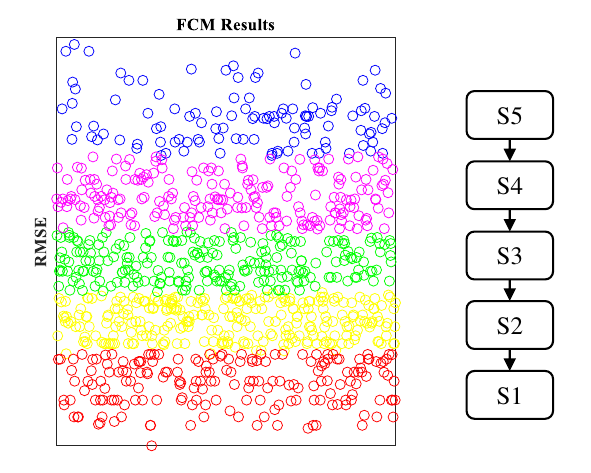}
	\caption{States using FCM clustering on the RMSE distribution.}
	\label{fig:FCM}
\end{figure}

\paragraph{Suboptimal Scheme}
Optimizing for multiple UAVs simultaneously can be complex. A practical, suboptimal approach is to define states based on localization performance. As shown in Fig.~\ref{fig:FCM}, we can define a finite number of states based on the distribution of the Root Mean Square Error (RMSE) using a clustering method like Fuzzy C-Means (FCM). The agent learns to select actions that move the system from poor states (e.g., S1) to good states (e.g., S5). The RL agent's goal is then to take actions that transition the system from a high-error state to a low-error state. 

The problem can be reformulated to find the minimum number of BSs, $R_s$, that can achieve a localization MSE below a certain threshold, $\mathcal{M_{ \max }}$, for all detected UAVs
\begin{align}
	\text{$\mathscr{P}_3$: } \text{min} &\quad R_s  \\
	\nonumber
	\text{s.t.} &\quad  \mathcal{M}_m(\mathbf{R},\mathbf{p}_t)<\mathcal{M_{ \max }} \text{ for all } m=1...M.
\end{align}
The Q-learning algorithm is then trained to satisfy this constraint. The state transitions from poor localization (high error) to accurate localization (error below $\mathcal{M_{ \max }}$), and the algorithm stops when the error constraint is met, yielding an efficient BS selection.

\section{Experimental and Simulation Results}

\subsection{Numerical Simulations}
To validate the performance of the proposed framework, we conduct a series of numerical simulations. We first describe the simulation setup and then present the results for collaborative localization and reinforcement learning-based BS selection.

\subsubsection{Simulation Settings}
\begin{figure}[t]
	\centering
	\includegraphics[width=0.85\columnwidth]{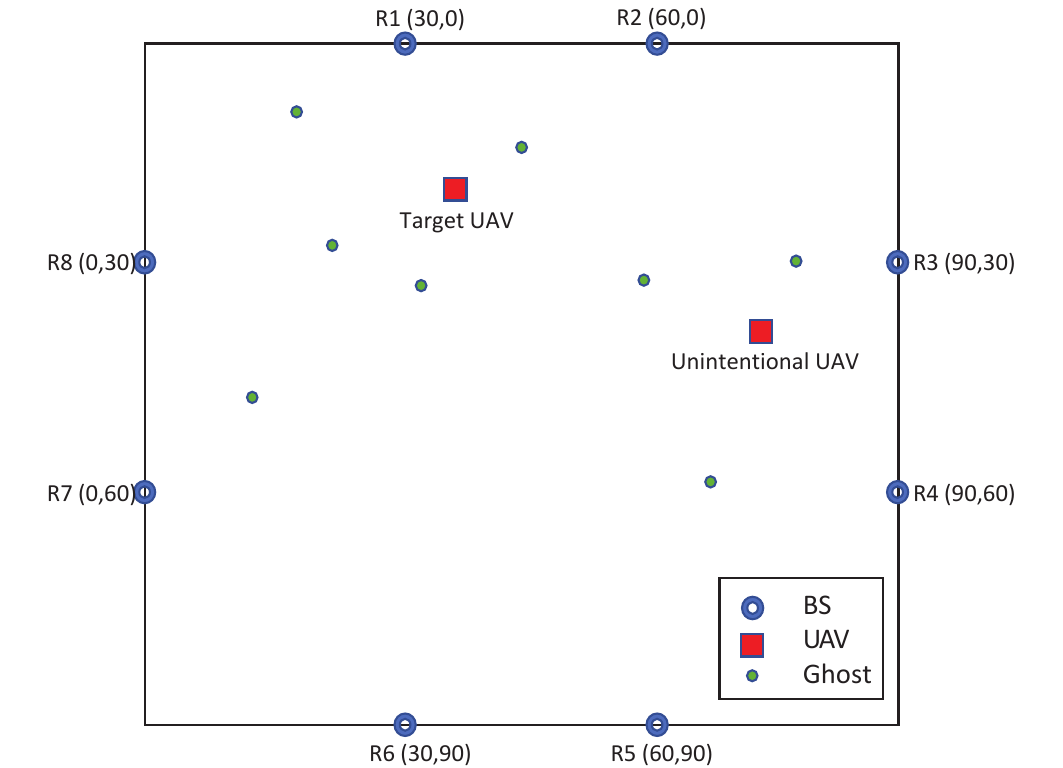}
	\caption{The simulation deployment.}
	\label{fig:Simulation_Scene}
\end{figure}

As shown in Fig.~\ref{fig:Simulation_Scene}, the simulation environment consists of a $90\,\text{m} \times 90\,\text{m}$ area with a grid resolution of $0.1\,\text{m}$. Two UAVs with different micro-Doppler characteristics are randomly located: ``target UAV'' (e.g., a quadcopter) and an ``unintentional UAV'' (e.g., a larger, helicopter-like UAV). 8 BSs are deployed at the edges of the map, each separated by $30\,\text{m}$. All BSs are assumed to be synchronized.

To simulate a complex environment, reflectors are randomly distributed according to a Poisson Point Process (PPP) with an expectation of 3, which generate multipath ghost echoes. The key simulation parameters for the BSs are listed in Tab.~\ref{tab:Radar_Simulation_Parameters}.

\begin{table}[h]
	\footnotesize
	\centering
	\caption{BS Simulation Parameters}
	\label{tab:Radar_Simulation_Parameters}
	\begin{tabular}{c|c|c|c}
		\hline
		\textbf{Parameter} & \textbf{Value} & \textbf{Parameter} & \textbf{Value} \\
		\hline
		Carrier Frequency (${f_0}$) & 24 GHz & Bandwidth ($B$) & 100 MHz \\
		Pulse Duration ($T$) & 1 ms & Samples ($L$) & 128 \\
		Number of Pulses ($M$) & 64 & ${P_{{\rm{FA}}}}$ & 0.001 \\
		\hline
	\end{tabular}
\end{table}

\subsubsection{Localization Results}
\begin{figure}[t]
	\centering
	\includegraphics[width=1\columnwidth]{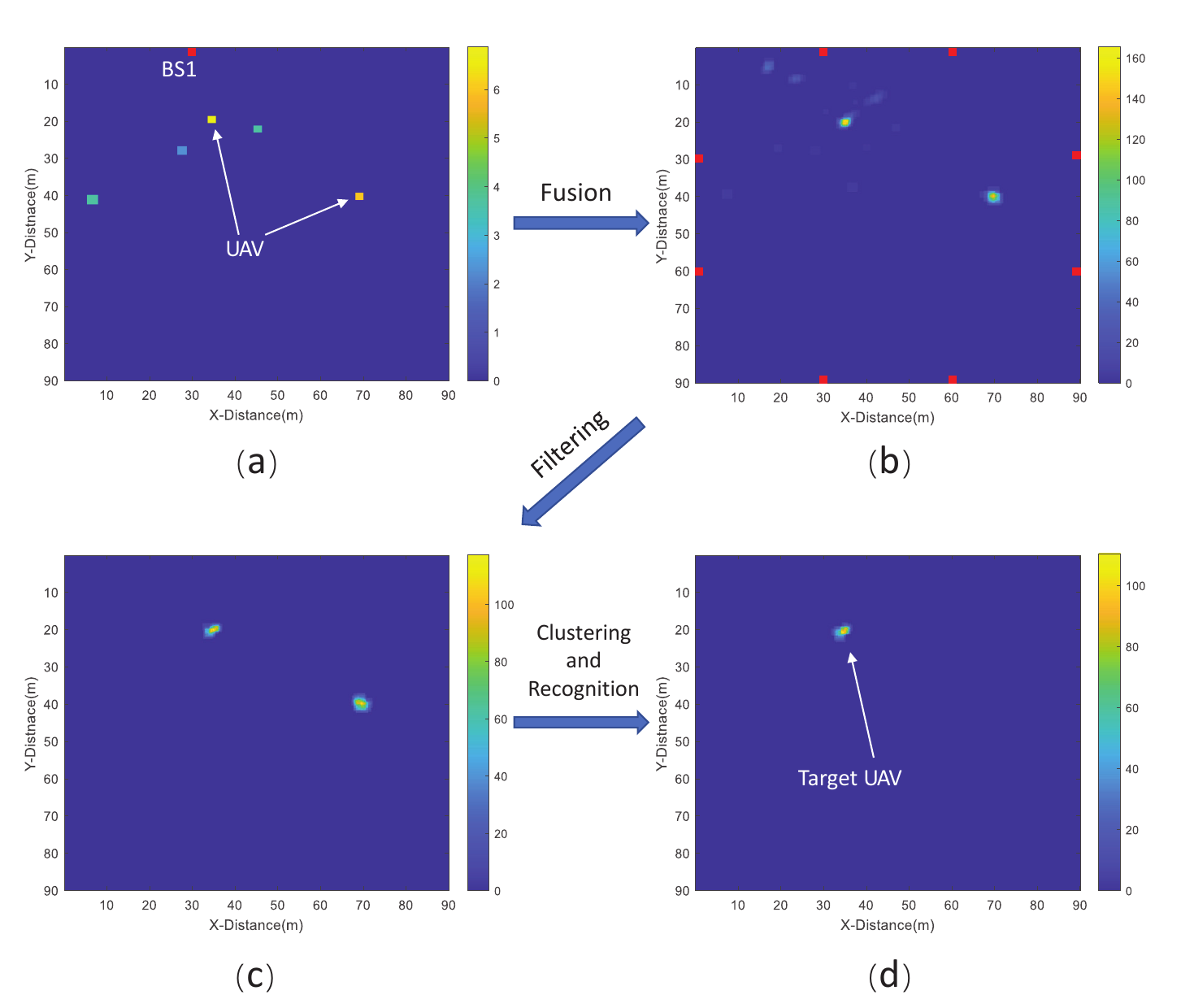}
	\caption{The step-by-step target positioning process.}
	\label{fig:simulation_fmwk}
\end{figure}

The simulated localization process is presented in Fig.~\ref{fig:simulation_fmwk}, where the color map represents the detection probability. The process demonstrates the progressive refinement from raw single-BS data to a precise, fused localization result.

\paragraph{Single BS Detection}
Initially, we evaluate the performance of a single BS (located at R1 in Fig.~\ref{fig:Simulation_Scene}). As shown in Fig.~\ref{fig:simulation_fmwk} (a), the single BS generates a probabilistic map containing multiple potential detections. It is unable to distinguish the true UAVs from the multipath ghosts, highlighting the challenge of localization in complex environments.

\paragraph{Observations from Multiple BSs}
Next, we apply the proposed data fusion framework to combine observations from all eight BSs. By comparing Fig.~\ref{fig:simulation_fmwk} (a) and Fig.~\ref{fig:simulation_fmwk}(b), it is clear that the fusion process significantly enhances the detection probabilities at the true UAV locations while suppressing the probabilities of the inconsistent ghost detections. By applying a threshold (set to half the peak probability value), the influence of multipath reflections is almost entirely mitigated, as shown in Fig.~\ref{fig:simulation_fmwk}(c). Finally, by applying the DBSCAN clustering and micro-Doppler recognition algorithms, the target of interest is successfully isolated and localized, as shown in Fig.~\ref{fig:simulation_fmwk} (d).

\begin{figure}[t]
	\centering
	\includegraphics[width=0.7\columnwidth]{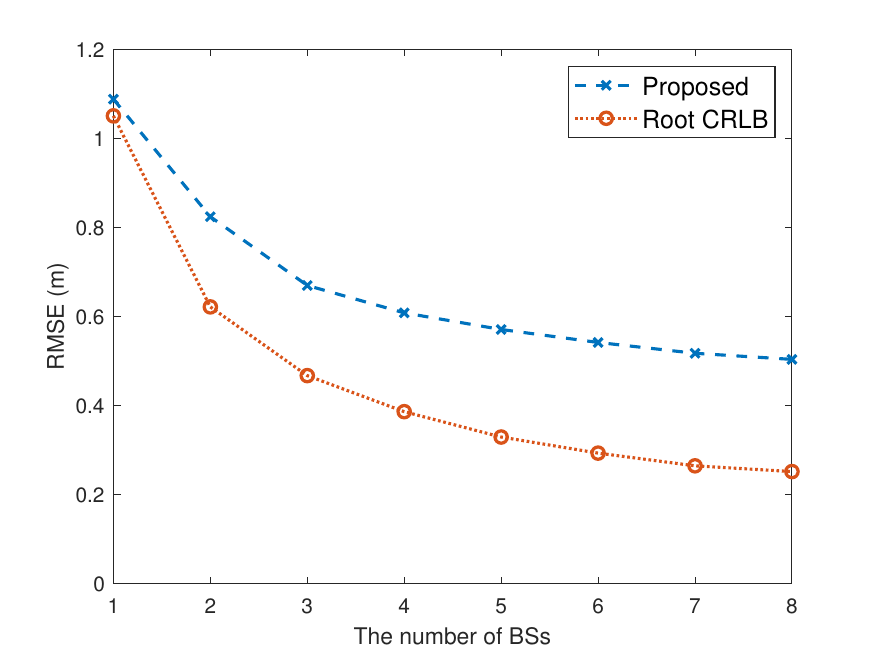}
	\caption{MSE versus the number of active BSs.}
	\label{fig:performance_curve}
\end{figure}

\begin{figure}[t]
	\centering
	\includegraphics[width=0.6\columnwidth]{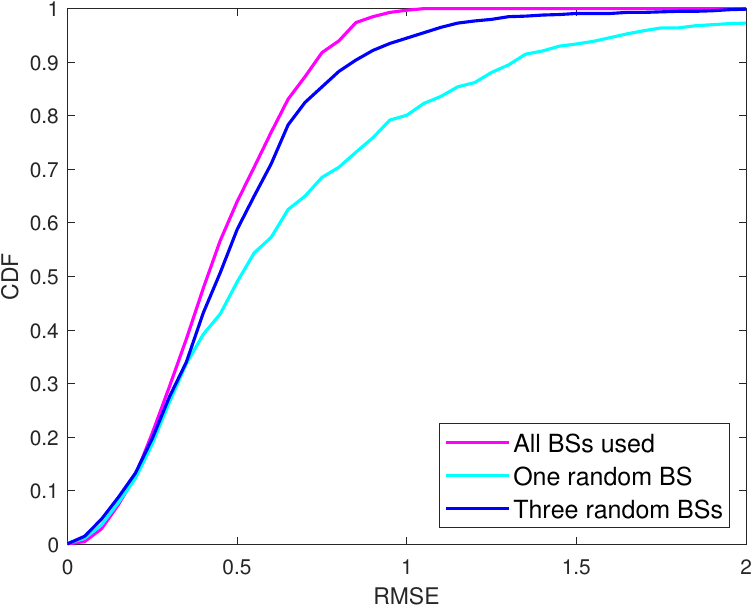}
	\caption{CDF of the positioning error.}
	\label{fig:cdf}
\end{figure}

\paragraph{Performance Analysis}
To quantitatively assess the performance, we conducted 5000 Monte Carlo simulations. Fig.~\ref{fig:performance_curve} shows the mean positioning error (MMSE) as a function of the number of active BSs, compared against the theoretical CRLB. The results clearly show that the localization error decreases as more BSs are used, demonstrating the benefit of collaborative sensing. The cumulative distribution function (CDF) of the positioning error, shown in Fig.~\ref{fig:cdf}, further illustrates that using all eight BSs provides a significant improvement over using a single or a small number of BSs.

\subsubsection{Node Selection Results}
We now evaluate the performance of the reinforcement learning-based BS selection algorithms, considering both the single-UAV objective ($\mathscr{P}_1$) and the multi-UAV objective ($\mathscr{P}_2$). We performed 2000 training iterations followed by 2000 Monte Carlo simulations, varying the number of UAVs from one to six.

\begin{figure}[t]
	\centering
	\includegraphics[width=0.6\columnwidth]{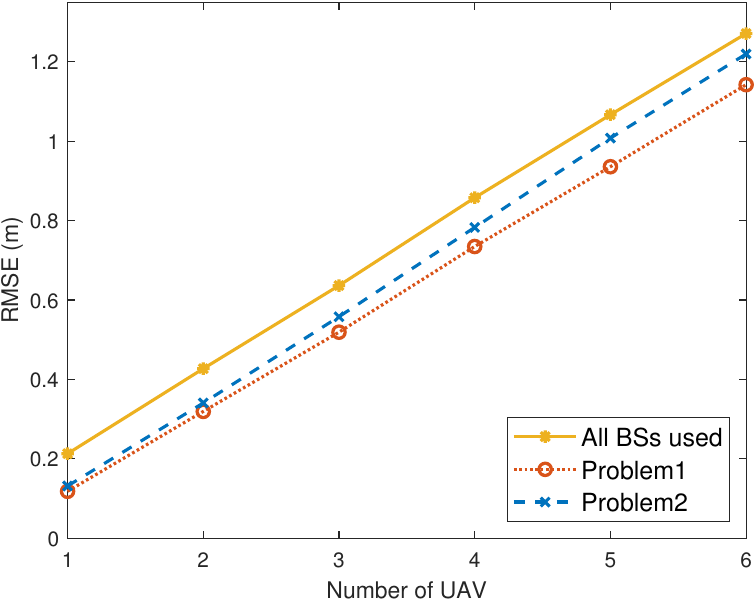}
	\caption{Positioning error versus the number of UAVs.}
	\label{fig:rmsevsnt}
\end{figure}

\begin{figure}[t]
	\centering
	\includegraphics[width=0.6\columnwidth]{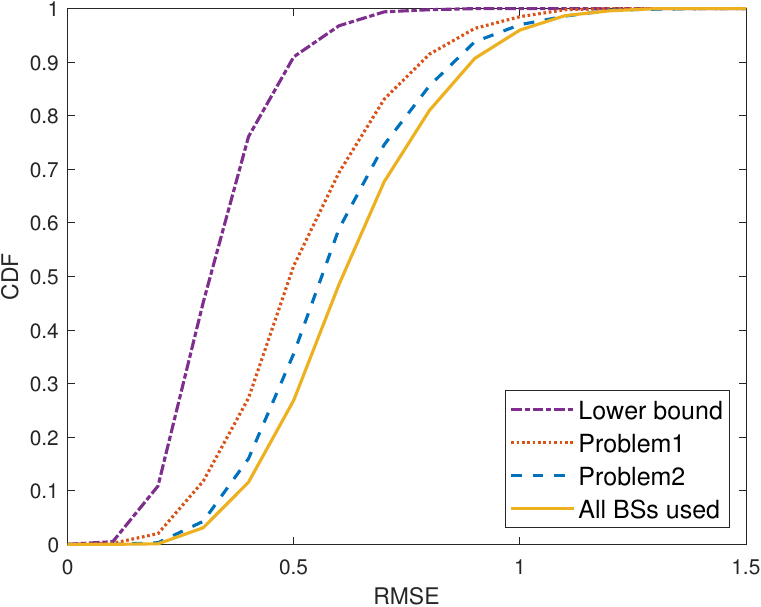}
	\caption{CDF of the positioning error for a three-UAV scenario.}
	\label{fig:cdf_error_3targets}
\end{figure}

\paragraph{RMSE Performance}
Fig.~\ref{fig:rmsevsnt} shows the total localization error versus the number of UAVs for three cases: using all BSs, using the $\mathscr{P}_1$ optimization, and using the $\mathscr{P}_2$ optimization. Both optimization methods outperform the baseline of using all BSs. The $\mathscr{P}_1$ method achieves the lowest error, while the $\mathscr{P}_2$ method  still offers a significant improvement of nearly 0.4 meters over the baseline. The CDF of the localization error for a three-UAV scenario is shown in Fig.~\ref{fig:cdf_error_3targets}, confirming that both optimization methods yield better performance than the baseline, though they remain above the theoretical lower bound due to the inclusion of resource cost in the reward function.

\begin{figure}[h]
	\footnotesize
	\centering
	\subfigure[Measuring equipment]{
		\label{SDR-kit 2400AD2}
		\includegraphics[width=0.4\columnwidth,height=0.3\columnwidth]{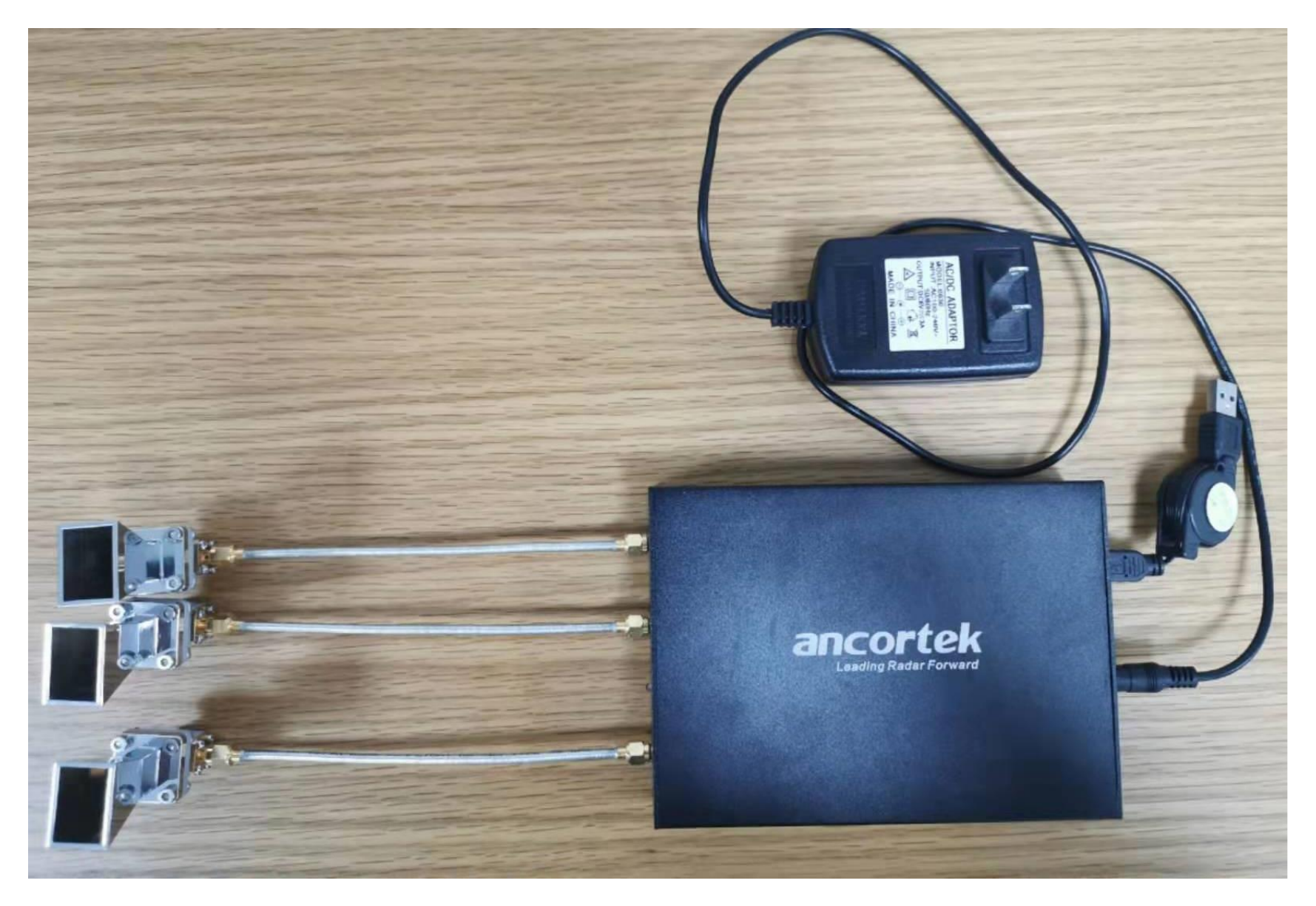}} 
	\subfigure[Experimental scene]{
		\label{fig:Experimental_Scene}
		\includegraphics[width=0.4\columnwidth]{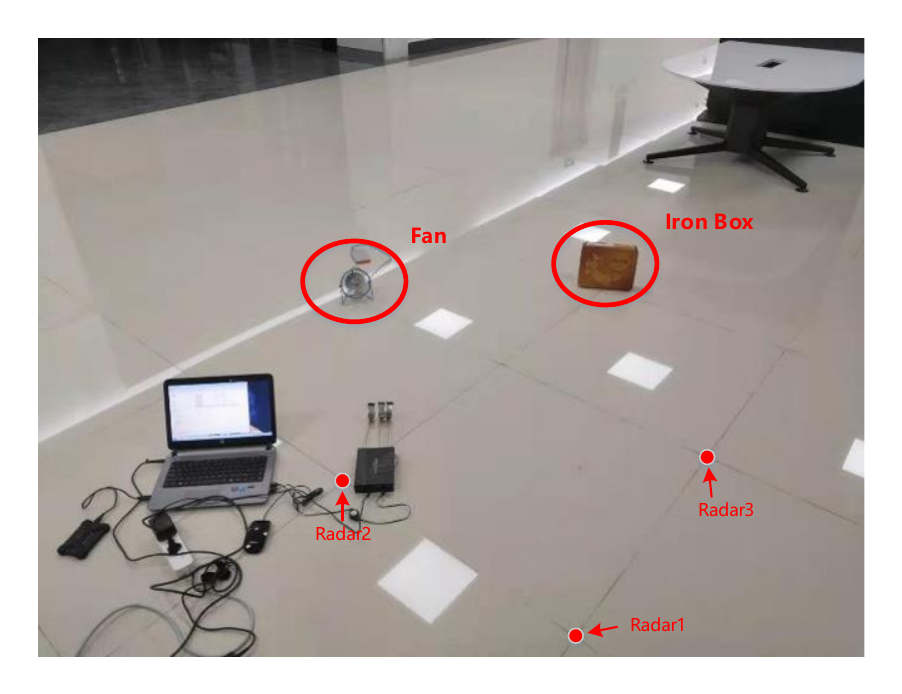}} 
	\caption{Settings of practical experiment.}
	\label{fig:Radar_and_Experimental_Scene}
\end{figure}
\begin{figure}[t]
	\centering
	\subfigure[MDS of fan (UAV). ] {\label{fig:Micro-doppler_Fan}\includegraphics[width=0.48\columnwidth]{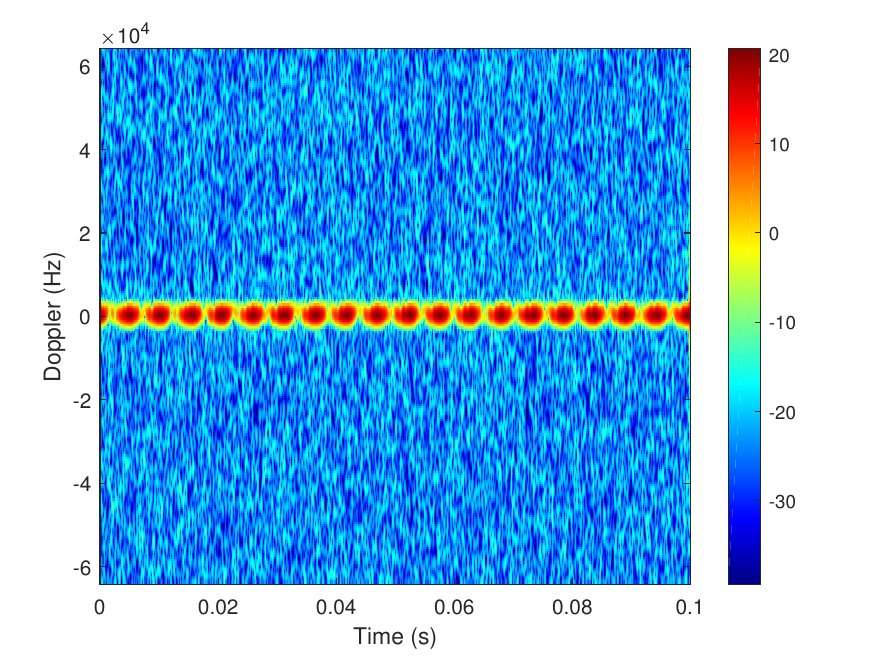}}
	\subfigure[MDS of box.]{\label{fig:Micro-doppler_Iron_Box}
		\includegraphics[width=0.48\columnwidth]{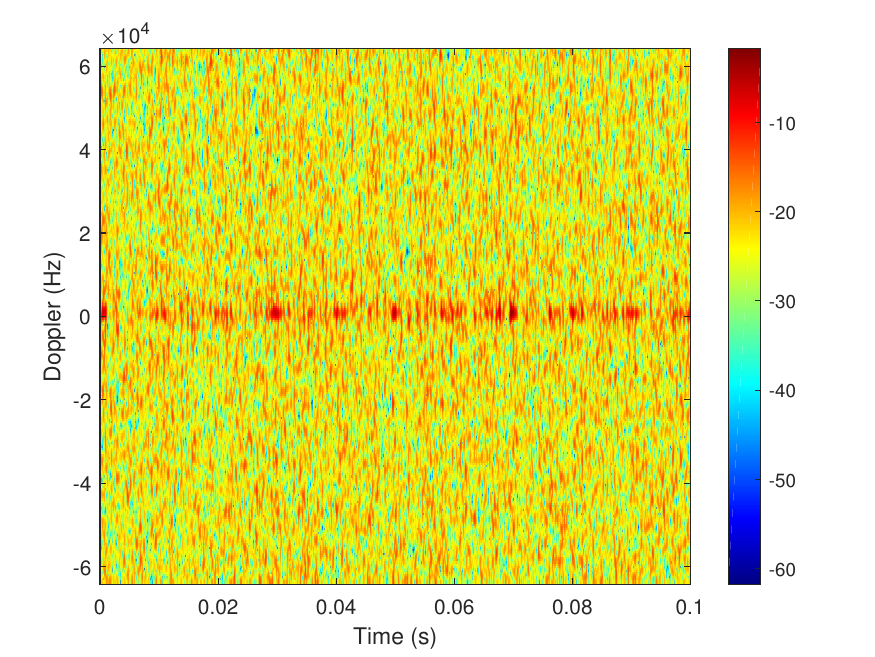}}
	\caption{Experimental micro-Doppler spectrograms.}
	\label{fig:Micro-doppler_Experiment}
\end{figure}

\subsection{Practical Experiments}
In addition to simulations, we conducted practical experiments to validate the proposed framework in a real-world indoor environment. The experiments demonstrate the effectiveness of both the micro-Doppler based UAV recognition and the multi-BS data fusion for improving localization accuracy.

\subsubsection{Experimental Setup}
The experiments were carried out using a K-band (24 GHz) FMCW radar kit (Ancortek SDR-kit 2400AD2), as shown in Fig.~\ref{fig:Radar_and_Experimental_Scene}(a). The radar is capable of generating the necessary waveforms for sensing and is equipped with one transmit and two receive horn antennas, allowing for both ToA and AoA measurements as described in Section IV-A.

The experimental scene, depicted in Fig.~\ref{fig:Radar_and_Experimental_Scene}(b), was set up in a cluttered indoor environment containing various stationary objects like chairs and desks. Two distinct objects were used for the experiment: a target of interest, represented by an electric fan to emulate the rotating blades of a UAV, and an unintentional object, a stationary iron box.

To simulate a multi-BS scenario, the radar was placed at three different locations sequentially. The data from each location was recorded and then processed as if it were collected simultaneously from three separate, spatially distributed BSs.

\subsubsection{Micro-Doppler Characterization}
First, we analyzed the micro-Doppler signatures of the fan (UAV) and the iron box to verify the recognition capability of our system. Figure.~\ref{fig:Micro-doppler_Experiment} shows the time-frequency spectrograms for both objects.

A clear difference is visible: the fan's rotating blades produce a distinct periodic micro-Doppler signature, whereas the iron box, being stationary, exhibits no such features. We applied the EMD-SVM classifier described in Section IV-B to these signals. The resulting recognition probabilities ($P_r$) are shown in Tab.~\ref{tab:Recognition_Probability_Experiment}. The classifier correctly identifies the fan as a UAV with high probability (0.8 to 0.93) across all three BS locations, while assigning a very low probability to the iron box. This confirms that the proposed feature enhancement method can effectively distinguish UAVs from other static or unintentional objects.

\begin{table}[h]
	\footnotesize
	\centering
	\caption{Experimental Target Recognition Probability ($P_r$)}
	\label{tab:Recognition_Probability_Experiment}
	\begin{tabular}{l|ccc}
		\hline
		\diagbox{\textbf{Object}}{\textbf{BS}} & \textbf{BS 1} & \textbf{BS 2} & \textbf{BS 3} \\
		\hline
		Fan (UAV) & 0.8 & 0.8667 & 0.9333 \\
		Iron Box & 0.2667 & 0 & 0.2 \\
		\hline
	\end{tabular}
\end{table}

\begin{figure}[h]
	\footnotesize
	\centering
	\subfigure[Detection fusion of 3 BSs.]{
		\label{labpd}
		\includegraphics[width=0.45\columnwidth]{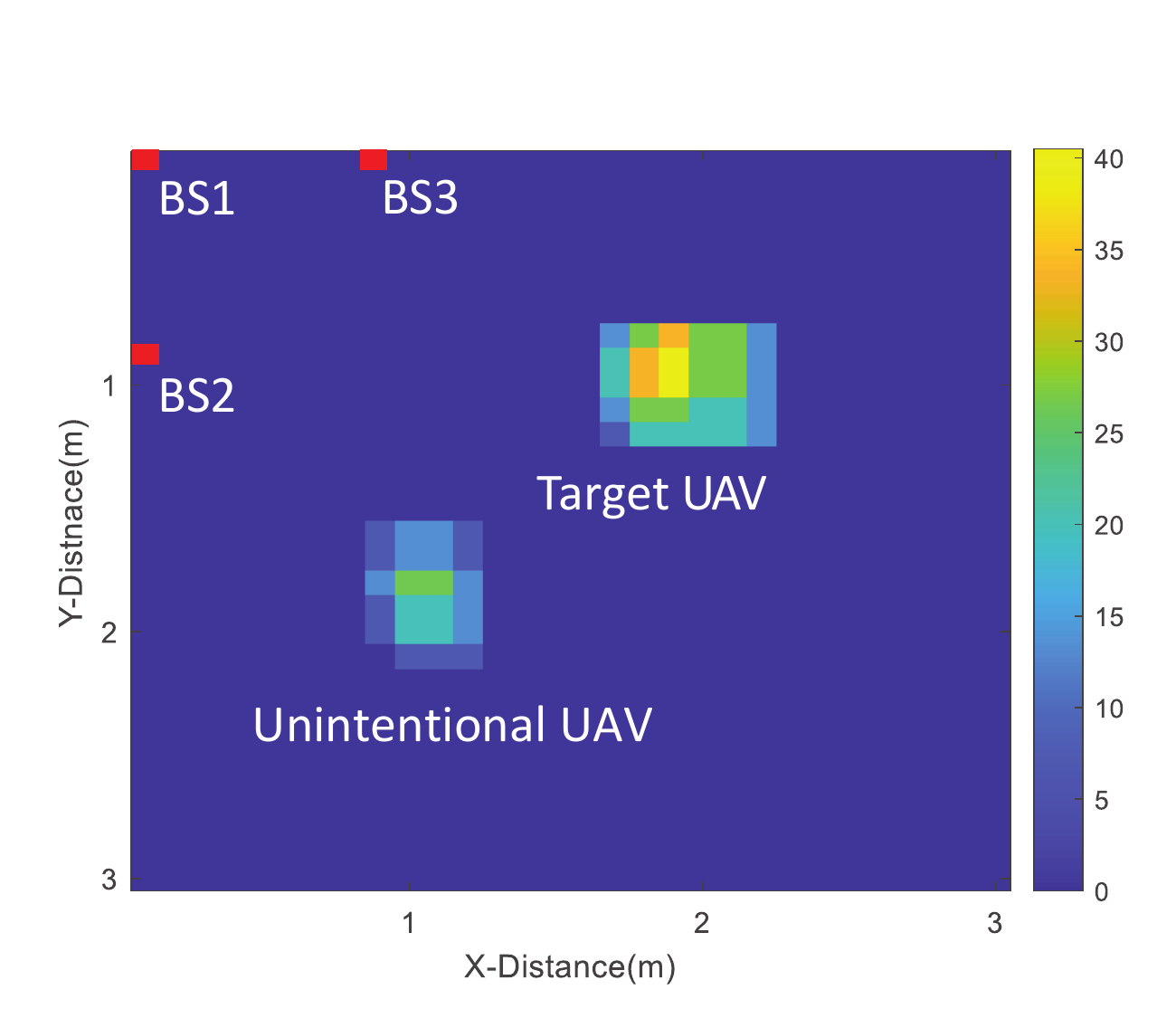}}
	\subfigure[Fusion after recognition.]{
		\label{labpr}
		\includegraphics[width=0.45\columnwidth]{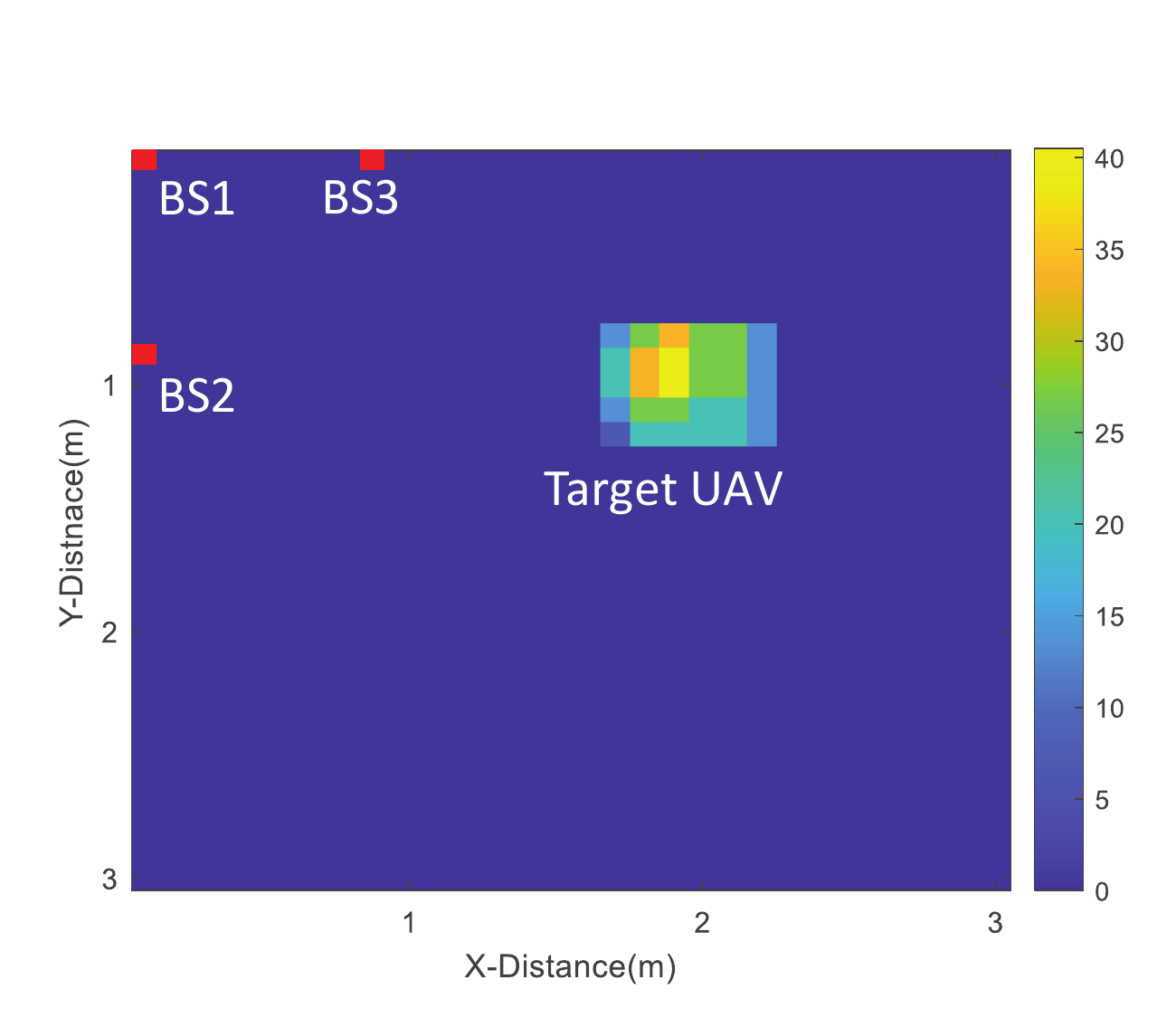}}
	\caption{Experimental fusion results.}
	\label{fig:fusion_lab}
\end{figure}

\subsubsection{Multi-BS Fusion and Localization}
Next, we evaluated the performance of the multi-BS data fusion framework. Figure \ref{fig:fusion_lab}(a) shows the fused detection map using the basic detection probability ($P_d$) from all three BS locations. While both the fan (target of interest) and the iron box (unintentional target) are detected, the map is noisy.

Figure \ref{fig:fusion_lab}(b) shows the result after applying our complete framework, where the fusion is weighted by the recognition probability ($P_r$). The framework successfully suppresses the detection of the iron box and isolates the fan, providing a clean and unambiguous localization of the UAV.

The positioning errors for individual BSs and various fused combinations are listed in Tab.~\ref{tab:Target_Positioning_Error}. The results show that fusing observations from multiple BSs provides a more "balanced" and often more accurate solution. For instance, while BS 2 has the lowest individual error, BS 1 and BS 3 have significant errors. Fusing all three BSs yields a final error of 0.0627 m, which is a substantial improvement over the average individual performance. This demonstrates that multi-BS observations provide spatial diversity that effectively mitigates multipath effects and improves overall detection accuracy and reliability.

\begin{table}[h]
	\footnotesize
	\centering
	\caption{Positioning Error of Experiments}
	\label{tab:Target_Positioning_Error}
	\begin{tabular}{l|c|c}
		\hline
		\diagbox{\textbf{BS Configuration}}{\textbf{Error (m)}} & \textbf{Error} & \textbf{Mean} \\
		\hline
		BS 1 & 0.0333 & \\
		BS 2 & 0.1414 & 0.1249 \\
		BS 3 & 0.2000 & \\
		\hline
		BS 1 and 2 & 0.0505 & \\
		BS 1 and 3 & 0.0514 & 0.0836 \\
		BS 2 and 3 & 0.1490 & \\
		\hline
		All three BSs & \multicolumn{2}{c}{0.0627} \\
		\hline
	\end{tabular}
\end{table}

\section{Conclusion}
In this paper, we investigated UAV sensing, detection and localization in the context of the emerging low-altitude economy, where the proliferation of LSS UAVs poses significant safety challenges. To address the limitations of conventional solutions, we proposed an RF-based cooperative sensing framework that leverages existing base stations, integrating CA-CFAR and MDS-based methods for robust UAV detection and recognition. A grid-based probabilistic data fusion and clustering strategy was introduced to mitigate ghost targets and improve multi-UAV localization accuracy. Furthermore, CRLB analysis and RL-based optimization were employed to balance localization performance and system efficiency. The feasibility and superiority of the proposed framework are verified through both numerical simulations and practical experiments. Our method is shown to reduce localization error by over $40\%$ through multi-station fusion, a result confirmed by experiments achieving centimeter-level accuracy, while our RL-based optimization efficiently manages network resources. Future work will extend the framework to more complex communication scenario and explore the co-design strategies between sensing and communication for large-scale UAV networks.

\section*{Appendix: Proof of Proposition 1}

The measurement equation for the position of a UAV, $\mathbf{p} = [x, y]^T$, as observed by ${N_{{\text{BS}}}}$ BSs, each with $K$ antennas, is given by the vector of observations $\mathbf{Z}$:
\begin{align}
	{\bf{Z}} = \left[ \begin{array}{c}
		r_{11} \\ \theta_{11} \\ \vdots \\ r_{nk} \\ \theta_{nk} \\ \vdots \\ r_{N_{BS}K} \\ \theta_{N_{BS}K}
	\end{array} \right] = \left[ \begin{array}{c}
		f_r(\mathbf{p}_{11}) \\ f_{\theta}(\mathbf{p}_{11}) \\ \vdots \\ f_r(\mathbf{p}_{nk}) \\ f_{\theta}(\mathbf{p}_{nk}) \\ \vdots \\ f_r(\mathbf{p}_{N_\text{BS}K}) \\ f_{\theta}(\mathbf{p}_{N_\text{BS}K})
	\end{array} \right] + {\bf{n}} = {\bf{H}}(\mathbf{p}) + {\bf{n}}
\end{align}
where $r_{nk}$ and $\theta_{nk}$ are the measured range and angle. The true range and angle, $f_r$ and $f_{\theta}$, are functions of the UAV's position $\mathbf{p}$ and the antenna's position $\mathbf{p}_{nk} = [x_{nk}, y_{nk}]^T$:
\begin{align}
	f_r(\mathbf{p}_{nk}) = \sqrt{(x_{nk} - x)^2 + (y_{nk} - y)^2}
\end{align}
\begin{align}
	f_{\theta}(\mathbf{p}_{nk}) = \arctan \left( \frac{y_{nk} - y}{x_{nk} - x} \right)
\end{align}
The measurement noise vector $\mathbf{n}$ is zero-mean Gaussian with covariance $\mathbf{\Lambda}$. The likelihood function is:
\begin{small}
	\begin{equation}
		\begin{array}{l}
			\rho(\mathbf{Z};\mathbf{p}) = C \cdot \exp \left\{ -\frac{1}{2}\sum_{n=1}^{N_\text{BS}}\sum_{k=1}^{K} \left[ \lambda_{nk}^{\text{TOA}}(\delta_{nk}^r)^2 + \lambda_{nk}^{\text{AOA}}(\delta_{nk}^{\theta})^2 \right] \right\}
		\end{array}
	\end{equation}
\end{small}
where $\delta_{nk}^r = r_{nk} - f_r(\mathbf{p}_{nk})$ and $\delta_{nk}^{\theta} = \theta_{nk} - f_{\theta}(\mathbf{p}_{nk})$ are the measurement errors, and $C$ represents the constant terms from the likelihood function's normalization factor, which will not influence the subsequent derivatives with respect to $\mathbf{p}$. The log-likelihood function is:
\begin{small}
	\begin{equation}
		\ln \rho(\mathbf{Z};\mathbf{p}) = \ln C - \frac{1}{2}\sum_{n=1}^{N_\text{BS}}\sum_{k=1}^{K} \left[ \lambda_{nk}^{\text{TOA}}(\delta_{nk}^r)^2 + \lambda_{nk}^{\text{AOA}}(\delta_{nk}^{\theta})^2 \right]
	\end{equation}
\end{small}
The first-order partial derivative of the log-likelihood is:
\begin{small}
	\begin{equation}
		\begin{split}
			\frac{\partial \ln \rho(\mathbf{Z} ; \mathbf{p})}{\partial \mathbf{p}} &= \sum_{n=1}^{N_{\text{BS}}} \sum_{k=1}^{K}\left[\lambda_{n k}^{\mathrm{TOA}}\left(\delta_{n k}^{r}\right) \frac{\partial f_{r}(\mathbf{p}_{nk})}{\partial \mathbf{p}} \right. \\
			&\quad \left. +\lambda_{n k}^{\mathrm{AOA}}\left(\delta_{n k}^{\theta}\right) \frac{\partial f_{\theta}(\mathbf{p}_{nk})}{\partial \mathbf{p}}\right]
		\end{split}
	\end{equation}
\end{small}
The second-order partial derivative is:
\begin{small}
	\begin{equation}
		\frac{\partial^2 \ln \rho(\mathbf{Z};\mathbf{p})}{\partial \mathbf{p} \partial \mathbf{p}^T} = \sum_{n=1}^{N_\text{BS}}\sum_{k=1}^{K} \left[ \lambda_{nk}^{\text{TOA}} \gamma_{r_{nk}} + \lambda_{nk}^{\text{AOA}} \gamma_{\theta_{nk}} \right]
	\end{equation}
\end{small}
where
\begin{small}
	\begin{equation}
		\gamma_{r_{nk}} = -\frac{\partial f_r}{\partial \mathbf{p}}\left(\frac{\partial f_r}{\partial \mathbf{p}}\right)^T + (\delta_{nk}^r)\frac{\partial^2 f_r}{\partial \mathbf{p} \partial \mathbf{p}^T}
	\end{equation}
\end{small}
\begin{small}
	\begin{equation}
		\gamma_{\theta_{nk}} = -\frac{\partial f_{\theta}}{\partial \mathbf{p}}\left(\frac{\partial f_{\theta}}{\partial \mathbf{p}}\right)^T + (\delta_{nk}^{\theta})\frac{\partial^2 f_{\theta}}{\partial \mathbf{p} \partial \mathbf{p}^T}
	\end{equation}
\end{small}
The Fisher Information Matrix (FIM) is the negative expectation of the second-order derivative. Since the expectation of the error terms $E[\delta_{nk}^r]$ and $E[\delta_{nk}^{\theta}]$ is zero, the second terms in $\gamma_{r_{nk}}$ and $\gamma_{\theta_{nk}}$ vanish.
\begin{small}
	\begin{equation}
		\begin{split}
			\mathbf{J}_e(\mathbf{p}) &= -E\left[ \frac{\partial^2 \ln \rho(\mathbf{Z};\mathbf{p})}{\partial \mathbf{p} \partial \mathbf{p}^T} \right] \\
			&= E \left[ \sum_{n=1}^{N_{\text{BS}}}\sum_{k=1}^{K} \left( \lambda_{nk}^{\text{TOA}} \frac{\partial f_r}{\partial \mathbf{p}} \frac{\partial f_r}{\partial \mathbf{p}^T} + \lambda_{nk}^{\text{AOA}} \frac{\partial f_{\theta}}{\partial \mathbf{p}} \frac{\partial f_{\theta}}{\partial \mathbf{p}^T} \right) \right]
		\end{split}
	\end{equation}
\end{small}
The partial derivatives of the range and angle functions are:
\begin{small}
	\begin{equation}
		\frac{\partial f_r}{\partial \mathbf{p}} = \left[ \begin{array}{c} -\cos\theta_{nk} \\ -\sin\theta_{nk} \end{array} \right] = -\mathbf{q}_{nk}^{\text{TOA}}
	\end{equation}
\end{small}
\begin{small}
	\begin{equation}
		\frac{\partial f_{\theta}}{\partial \mathbf{p}} = \frac{1}{f_r} \left[ \begin{array}{c} \sin\theta_{nk} \\ -\cos\theta_{nk} \end{array} \right] = \frac{1}{f_r} \mathbf{q}_{nk}^{\text{AOA}}
	\end{equation}
\end{small}
The outer product for the ToA component is:
\begin{small}
	\begin{equation}
		\begin{array}{l}
			\frac{\partial f_r}{\partial \mathbf{p}} \left( \frac{\partial f_r}{\partial \mathbf{p}} \right)^T = \left[ \begin{array}{c}
				-\cos\theta_{nk}\\
				-\sin\theta_{nk}
			\end{array} \right]\left[ {-\cos\theta_{nk}} \;\; {-\sin\theta_{nk}} \right]\\
			\qquad = \left[ \begin{array}{cc}
				\cos^2\theta_{nk} & \cos\theta_{nk}\sin\theta_{nk}\\
				\cos\theta_{nk}\sin\theta_{nk} & \sin^2\theta_{nk}
			\end{array} \right]\\
			\qquad = \mathbf{q}_{nk}^{\text{TOA}}(\mathbf{q}_{nk}^{\text{TOA}})^T
		\end{array}
	\end{equation}
\end{small}
And the outer product for the AoA component is:
\begin{small}
	\begin{equation}
		\begin{split}
			\frac{\partial f_{\theta}}{\partial \mathbf{p}} \left( \frac{\partial f_{\theta}}{\partial \mathbf{p}} \right)^T &= \frac{1}{f_r^2} \left[ \begin{array}{c}
				\sin\theta_{nk}\\
				-\cos\theta_{nk}
			\end{array} \right]\left[ {\sin\theta_{nk}} \;\; {-\cos\theta_{nk}} \right]\\
			& = \frac{1}{f_r^2} \left[ \begin{array}{cc}
				\sin^2\theta_{nk} & -\sin\theta_{nk}\cos\theta_{nk}\\
				-\sin\theta_{nk}\cos\theta_{nk} & \cos^2\theta_{nk}
			\end{array} \right]\\
			& = \frac{1}{f_r^2} \mathbf{q}_{nk}^{\text{AOA}}(\mathbf{q}_{nk}^{\text{AOA}})^T
		\end{split}
	\end{equation}
\end{small}
Substituting these results back into the FIM expression yields the final form:
\begin{small}
	\begin{equation}
		\begin{split}
			\mathbf{J}_{\mathrm{e}}(\mathbf{p})
			&=\sum_{n=1}^{N_{\text{BS}}} \sum_{k=1}^{K}\left\{ \lambda_{nk}^{\mathrm{TOA}} (\mathbf{q}_{nk}^{\mathrm{TOA}})(\mathbf{q}_{nk}^{\mathrm{TOA}})^T \right. \\
			& \quad \left. + \frac{\lambda_{nk}^{\mathrm{AOA}}}{f_r^2} (\mathbf{q}_{nk}^{\mathrm{AOA}})(\mathbf{q}_{nk}^{\mathrm{AOA}})^T \right\} \\
			&= \sum_{n=1}^{N_{\text{BS}}} \sum_{k=1}^K \left( \mathbf{J}_{nk}^{\text{TOA}}(\mathbf{p}) + \mathbf{J}_{nk}^{\text{AOA}}(\mathbf{p}) \right)
		\end{split}
	\end{equation}
\end{small}

\bibliographystyle{IEEETran}
\bibliography{Radar_Fusion_References}

\begin{thebibliography}{10}
\providecommand{\url}[1]{#1}
\csname url@samestyle\endcsname
\providecommand{\newblock}{\relax}
\providecommand{\bibinfo}[2]{#2}
\providecommand{\BIBentrySTDinterwordspacing}{\spaceskip=0pt\relax}
\providecommand{\BIBentryALTinterwordstretchfactor}{4}
\providecommand{\BIBentryALTinterwordspacing}{\spaceskip=\fontdimen2\font plus
\BIBentryALTinterwordstretchfactor\fontdimen3\font minus
  \fontdimen4\font\relax}
\providecommand{\BIBforeignlanguage}[2]{{%
\expandafter\ifx\csname l@#1\endcsname\relax
\typeout{** WARNING: IEEEtran.bst: No hyphenation pattern has been}%
\typeout{** loaded for the language `#1'. Using the pattern for}%
\typeout{** the default language instead.}%
\else
\language=\csname l@#1\endcsname
\fi
#2}}
\providecommand{\BIBdecl}{\relax}
\BIBdecl

\bibitem{huang2024potential}
H.~Huang, J.~Su, and F.-Y. Wang, ``The potential of low-altitude airspace: The
  future of urban air transportation,'' \emph{IEEE Transactions on Intelligent
  Vehicles}, 2024.

\bibitem{he2025ubiquitous}
D.~He, W.~Yuan, J.~Wu, and R.~Liu, ``Ubiquitous {UAV} communication enabled
  low-altitude economy: Applications, techniques, and 3gpp's efforts,''
  \emph{IEEE Network}, 2025.

\bibitem{ping2025multimodal}
Y.~Ping, T.~Liang, H.~Ding, G.~Lei, J.~Wu, X.~Zou, K.~Shi, R.~Shao, C.~Zhang,
  W.~Zhang \emph{et~al.}, ``Multimodal large language models-enabled {UAV}
  swarm: Towards efficient and intelligent autonomous aerial systems,''
  \emph{arXiv preprint arXiv:2506.12710}, 2025.

\bibitem{wu2021comprehensive}
Q.~Wu, J.~Xu, Y.~Zeng, D.~W.~K. Ng, N.~Al-Dhahir, R.~Schober, and A.~L.
  Swindlehurst, ``A comprehensive overview on {5G}-and-beyond networks with
  {UAVs}: From communications to sensing and intelligence,'' \emph{IEEE Journal
  on Selected Areas in Communications}, vol.~39, no.~10, pp. 2912--2945, 2021.

\bibitem{LTHNETWORK}
T.~Liang, T.~Zhang, and Q.~Zhang, ``Toward seamless localization and
  communication: A {Satellite-UAV} {NTN} architecture,'' \emph{IEEE Network},
  vol.~38, no.~4, pp. 103--110, 2024.

\bibitem{lth2022icc}
T.~Liang, W.~Liu, J.~Yang, and T.~Zhang, ``Age of information based scheduling
  for {UAV} aided emergency communication networks,'' in \emph{ICC 2022 - IEEE
  International Conference on Communications}, 2022, pp. 5128--5133.

\bibitem{lth2024twc}
T.~Liang, T.~Zhang, Q.~Wu, W.~Liu, D.~Li, Z.~Xie, D.~Li, and Q.~Zhang, ``Age of
  information based scheduling for {UAV} aided localization and
  communication,'' \emph{IEEE Transactions on Wireless Communications},
  vol.~23, no.~5, pp. 4610--4626, 2024.

\bibitem{pothana2023uas}
P.~Pothana, J.~Joy, P.~Snyder, and S.~Vidhyadharan, ``Uas air-risk assessment
  in and around airports,'' in \emph{2023 Integrated Communication, Navigation
  and Surveillance Conference (ICNS)}, 2023, pp. 1--11.

\bibitem{yang2024access}
J.~Yang, Y.~Wang, X.~Hang, and D.~Delahaye, ``A review on airspace design and
  risk assessment for urban air mobility,'' \emph{IEEE Access}, vol.~12, pp.
  157\,599--157\,611, 2024.

\bibitem{yuan2020wcl}
W.~Yuan, C.~Liu, F.~Liu, S.~Li, and D.~W.~K. Ng, ``Learning-based predictive
  beamforming for {UAV} communications with jittering,'' \emph{IEEE Wireless
  Communications Letters}, vol.~9, no.~11, pp. 1970--1974, 2020.

\bibitem{jin2025advancing}
H.~Jin, W.~Yuan, J.~Wu, J.~Wang, D.~Niyato, X.~Wang, G.~K. Karagiannidis,
  Z.~Lin, Y.~Gong, D.~I. Kim \emph{et~al.}, ``Advancing the control of
  low-altitude wireless networks: Architecture, design principles, and future
  directions,'' \emph{arXiv preprint arXiv:2508.07967}, 2025.

\bibitem{wang2025twc}
Y.~Wang, K.~Zu, L.~Xiang, Q.~Zhang, Z.~Feng, J.~Hu, and K.~Yang, ``Isac enabled
  cooperative detection for cellular-connected {UAV} network,'' \emph{IEEE
  Transactions on Wireless Communications}, vol.~24, no.~2, pp. 1541--1554,
  2025.

\bibitem{wu2025toward}
J.~Wu, W.~Yuan, Q.~Cheng, and H.~Jin, ``Toward dual-functional {LAWN}:
  Control-aware system design for aerodynamics-aided {UAV} formations,''
  \emph{arXiv preprint arXiv:2507.19910}, 2025.

\bibitem{cohen2021urban}
A.~P. Cohen, S.~A. Shaheen, and E.~M. Farrar, ``Urban air mobility: History,
  ecosystem, market potential, and challenges,'' \emph{IEEE Transactions on
  Intelligent Transportation Systems}, vol.~22, no.~9, pp. 6074--6087, 2021.

\bibitem{dong2025iotj}
C.~Dong, Y.~Wang, W.~Wang, X.~Zhu, M.~Zhang, and Q.~Wu, ``Low-altitude centric
  space-air-ground integrated network: Evolutions, challenges and
  countermeasures,'' \emph{IEEE Internet of Things Journal}, pp. 1--1, 2025.

\bibitem{huang2021icca}
X.~Huang, Z.~Wang, Q.~Peng, H.~Xu, and Z.~He, ``{LSS UAV} target intelligent
  detection in urban complex environment,'' in \emph{2021 IEEE 3rd
  International Conference on Civil Aviation Safety and Information Technology
  (ICCASIT)}, 2021, pp. 648--650.

\bibitem{lei2025enhancing}
G.~Lei, T.~Liang, Y.~Ping, X.~Chen, L.~Zhou, J.~Wu, X.~Zhang, H.~Ding,
  X.~Zhang, W.~Yuan \emph{et~al.}, ``Enhancing low-altitude airspace security:
  Mllm-enabled {UAV} intent recognition,'' \emph{arXiv preprint
  arXiv:2509.06312}, 2025.

\bibitem{li2025iotj}
Z.~Li, Z.~Gao, K.~Wang, Y.~Mei, C.~Zhu, L.~Chen, X.~Wu, and D.~Niyato,
  ``Unauthorized {UAV} countermeasure for low-altitude economy: Joint
  communications and jamming based on {MIMO} cellular systems,'' \emph{IEEE
  Internet of Things Journal}, vol.~12, no.~6, pp. 6659--6672, 2025.

\bibitem{liu2025mape}
M.~Liu, H.~Xiao, G.~Pan, J.~Zhou, W.~Li, and X.~Xi, ``How to achieve
  large-scale development in the low-altitude economy,'' in \emph{2024 IEEE
  10th International Symposium on Microwave, Antenna, Propagation and EMC
  Technologies for Wireless Communications (MAPE)}, 2024, pp. 1--4.

\bibitem{lth2024jsas}
T.~Liang, T.~Zhang, S.~Zhou, W.~Liu, D.~Li, and Q.~Zhang, ``{UAV}-aided
  localization and communication: Joint frame structure, beamwidth, and power
  allocation,'' \emph{IEEE Journal of Selected Areas in Sensors}, vol.~1, pp.
  154--165, 2024.

\bibitem{lth2025iotj}
T.~Liang, H.~Ding, Y.~Ping, T.~Zhang, L.~Zhou, Q.~Zhang, and T.~Q.~S. Quek,
  ``Satellite-assisted uav control: Sensing and communication scheduling for
  energy efficient data collection,'' \emph{IEEE Internet of Things Journal},
  pp. 1--1, 2025.

\bibitem{niu2021icba}
R.~Niu, Y.~Qu, and Z.~Wang, ``{UAV} detection based on improved yolov4 object
  detection model,'' in \emph{2021 2nd International Conference on Big Data \&
  Artificial Intelligence \& Software Engineering (ICBASE)}, 2021, pp. 25--29.

\bibitem{wang2024spl}
Z.~Wang, Z.~Cao, J.~Xie, W.~Zhang, and Z.~He, ``Rf-based drone detection
  enhancement via a generalized denoising and interference-removal framework,''
  \emph{IEEE Signal Processing Letters}, vol.~31, pp. 929--933, 2024.

\bibitem{ezuma2020ojcs}
M.~Ezuma, F.~Erden, C.~Kumar~Anjinappa, O.~Ozdemir, and I.~Guvenc, ``Detection
  and classification of {UAVs} using rf fingerprints in the presence of wi-fi
  and bluetooth interference,'' \emph{IEEE Open Journal of the Communications
  Society}, vol.~1, pp. 60--76, 2020.

\bibitem{xie2024iotj}
W.~Xie, Y.~Wan, G.~Wu, Y.~Li, F.~Zhou, and Q.~Wu, ``An {RF-Visual} directional
  fusion framework for precise {UAV} positioning,'' \emph{IEEE Internet of
  Things Journal}, vol.~11, no.~22, pp. 36\,736--36\,747, 2024.

\bibitem{yuan2021integrated}
W.~Yuan, Z.~Wei, S.~Li, J.~Yuan, and D.~W.~K. Ng, ``Integrated sensing and
  communication-assisted orthogonal time frequency space transmission for
  vehicular networks,'' \emph{IEEE Journal of Selected Topics in Signal
  Processing}, vol.~15, no.~6, pp. 1515--1528, 2021.

\bibitem{JiaYanZha:C22}
M.~Jia, J.~Yang, and T.~Zhang, ``Power allocation in infrastructure limited
  integration sensing and localization wireless networks,'' in \emph{2022 IEEE
  96th Vehicular Technology Conference (VTC2022-Fall)}, 2022, pp. 1--5.

\bibitem{liu2020radar}
F.~Liu, W.~Yuan, C.~Masouros, and J.~Yuan, ``Radar-assisted predictive
  beamforming for vehicular links: Communication served by sensing,''
  \emph{IEEE Transactions on Wireless Communications}, vol.~19, no.~11, pp.
  7704--7719, 2020.

\bibitem{liu2022integrated}
F.~Liu, Y.~Cui, C.~Masouros, J.~Xu, T.~X. Han, Y.~C. Eldar, and S.~Buzzi,
  ``Integrated sensing and communications: Toward dual-functional wireless
  networks for {6G} and beyond,'' \emph{IEEE journal on selected areas in
  communications}, vol.~40, no.~6, pp. 1728--1767, 2022.

\bibitem{haonan2023pimrc}
H.~He, T.~Liang, and T.~Zhang, ``Robust beamforming for isac systems in highly
  dynamic scenarios,'' in \emph{2023 IEEE 34th Annual International Symposium
  on Personal, Indoor and Mobile Radio Communications (PIMRC)}, 2023, pp. 1--6.

\bibitem{zhang2024predictive}
X.~Zhang, W.~Yuan, C.~Liu, J.~Wu, and D.~W.~K. Ng, ``Predictive beamforming for
  vehicles with complex behaviors in {ISAC} systems: A deep learning
  approach,'' \emph{IEEE Journal of Selected Topics in Signal Processing},
  vol.~18, no.~5, pp. 828--841, 2024.

\bibitem{yan2025deep}
Z.~Yan, W.~Yuan, X.~Zhang, C.~Liu, J.~Wu, and T.~Q. Quek, ``Deep learning-based
  compensation mechanism for {UAV} sensing via {OTFS} signaling,'' \emph{IEEE
  Internet of Things Journal}, 2025.

\bibitem{wu2025maga}
J.~Wu, W.~Yuan, X.~Zhang, Y.~Yu, Y.~Cui, F.~Liu, G.~Sun, J.~Wang, D.~Niyato,
  and D.~I. Kim, ``Toward multi-functional {LAWNs} with {ISAC}: Opportunities,
  challenges, and the road ahead,'' \emph{arXiv preprint arXiv:2508.17354},
  2025.

\bibitem{wu2023uavs}
J.~Wu, W.~Yuan, and L.~Hanzo, ``When {UAVs} meet {ISAC}: Real-time trajectory
  design for secure communications,'' \emph{IEEE Transactions on Vehicular
  Technology}, vol.~72, no.~12, pp. 16\,766--16\,771, 2023.

\bibitem{zhao2023tgrs}
Y.~Zhao and Y.~Su, ``Estimation of micro-doppler parameters with combined null
  space pursuit methods for the identification of {LSS UAVs},'' \emph{IEEE
  Transactions on Geoscience and Remote Sensing}, vol.~61, pp. 1--11, 2023.

\bibitem{zhu2022tpami}
P.~Zhu, L.~Wen, D.~Du, X.~Bian, H.~Fan, Q.~Hu, and H.~Ling, ``Detection and
  tracking meet drones challenge,'' \emph{IEEE Transactions on Pattern Analysis
  and Machine Intelligence}, vol.~44, no.~11, pp. 7380--7399, 2022.

\bibitem{wang2025sensor}
Q.~Wang, X.~Wei, and M.~Gao, ``Lssdnet: A low-slow-small target detection
  method based on centimeter wave radar in the urban environment,'' \emph{IEEE
  Sensors Journal}, vol.~25, no.~12, pp. 22\,118--22\,137, 2025.

\bibitem{selvi2025wisp}
A.~A. Selvi and S.~P.~K. Babu, ``Design and simulation of an {FMCW/CW} {X-Band}
  radar for detection/classification of {LSS} targets,'' in \emph{2025
  International Conference on Wireless Communications Signal Processing and
  Networking (WiSPNET)}, 2025, pp. 1--6.

\bibitem{yuan2025sensor}
W.~Yuan, X.~Chen, X.~Du, J.~Guan, J.~Wang, and T.~Lan, ``A low slow small
  target classification network model based on k-band radar dynamic
  multifeature data fusion,'' \emph{IEEE Sensors Journal}, vol.~25, no.~1, pp.
  1656--1668, 2025.

\bibitem{ahir2024map}
Y.~Ahirrao, R.~P. Yadav, and S.~Kumar, ``Real-time {UAV} detection through {RF}
  signal analysis and machine learning,'' in \emph{2024 IEEE Microwaves,
  Antennas, and Propagation Conference (MAPCON)}, 2024, pp. 1--4.

\bibitem{ak2024tvt}
R.~Akter, V.-S. Doan, A.~Zainudin, and D.-S. Kim, ``Sparsely connected low
  complexity {CNN} for unmanned vehicles detection-sensing {RF} signal,''
  \emph{IEEE Transactions on Vehicular Technology}, vol.~73, no.~10, pp.
  14\,236--14\,251, 2024.

\bibitem{GaoJiaZha:C21}
B.~Gao, M.~Jia, T.~Zhang, and Q.~Zhang, ``Reliable target positioning in
  complicated environments using multiple radar observations,'' in \emph{2021
  IEEE Global Communications Conference (GLOBECOM)}, 2021, pp. 1--6.

\bibitem{ghostgeometrypaper}
R.~Feng, E.~De~Greef, M.~Rykunov, H.~Sahli, S.~Pollin, and A.~Bourdoux,
  ``Multipath ghost recognition for indoor {MIMO} radar,'' \emph{IEEE
  Transactions on Geoscience and Remote Sensing}, pp. 1--10, 2021.

\bibitem{wanJiaMen:C22}
X.~Wang, M.~Jia, X.~Meng, and T.~Zhang, ``Multipath ghosts mitigation for
  radar-based positioning systems,'' in \emph{2022 IEEE 96th Vehicular
  Technology Conference (VTC2022-Fall)}, 2022, pp. 1--6.

\bibitem{ghostdataset}
F.~Kraus, N.~Scheiner, W.~Ritter, and K.~Dietmayer, ``The radar ghost dataset:
  An evaluation of ghost objects in automotive radar data,'' in \emph{2021
  IEEE/RSJ International Conference on Intelligent Robots and Systems (IROS)},
  2021, pp. 8570--8577.

\bibitem{ghoz2021wim}
D.~Ghozlani, A.~Omri, S.~Bouallegue, H.~Chamkhia, and R.~Bouallegue,
  ``Stochastic geometry-based analysis of joint radar and communication-enabled
  cooperative detection systems,'' in \emph{2021 17th International Conference
  on Wireless and Mobile Computing, Networking and Communications (WiMob)},
  2021, pp. 325--330.

\bibitem{jiayan2022iotj}
J.~Yang, T.~Zhang, X.~Wu, T.~Liang, and Q.~Zhang, ``Efficient scheduling in
  space-air-ground-integrated localization networks,'' \emph{IEEE Internet of
  Things Journal}, vol.~9, no.~18, pp. 17\,689--17\,704, 2022.

\bibitem{jin2021access}
G.~Ji, C.~Song, and H.~Huo, ``Detection and identification of low-slow-small
  rotor unmanned aerial vehicle using micro-doppler information,'' \emph{IEEE
  Access}, vol.~9, pp. 99\,995--100\,008, 2021.

\bibitem{sun2021tgrs}
Y.~Sun, S.~Abeywickrama, L.~Jayasinghe, C.~Yuen, J.~Chen, and M.~Zhang,
  ``Micro-doppler signature-based detection, classification, and localization
  of small uav with long short-term memory neural network,'' \emph{IEEE
  Transactions on Geoscience and Remote Sensing}, vol.~59, no.~8, pp.
  6285--6300, 2021.

\bibitem{ztt2016twc}
T.~Zhang, A.~F. Molisch, Y.~Shen, Q.~Zhang, H.~Feng, and M.~Z. Win, ``Joint
  power and bandwidth allocation in wireless cooperative localization
  networks,'' \emph{IEEE Transactions on Wireless Communications}, vol.~15,
  no.~10, pp. 6527--6540, 2016.

\bibitem{Bek2006TSP}
I.~Bekkerman and J.~Tabrikian, ``Target detection and localization using mimo
  radars and sonars,'' \emph{IEEE Transactions on Signal Processing}, vol.~54,
  no.~10, pp. 3873--3883, 2006.

\bibitem{yuan2016tvt}
W.~Yuan, N.~Wu, B.~Etzlinger, H.~Wang, and J.~Kuang, ``Cooperative joint
  localization and clock synchronization based on gaussian message passing in
  asynchronous wireless networks,'' \emph{IEEE Transactions on Vehicular
  Technology}, vol.~65, no.~9, pp. 7258--7273, 2016.

\bibitem{ref6}
E.~M. Mohamed and M.~M. Fouda, ``{OTFS-Based Proactive Dynamic UAV Positioning
  for High-Speed Train Coverage},'' \emph{IEEE Open Journal of the
  Communications Society}, vol.~5, pp. 5718--5734, 2024.

\bibitem{zafar2019rss}
F.~Zafari, A.~Gkelias, and K.~K. Leung, ``A survey of indoor localization
  systems and technologies,'' \emph{IEEE Communications Surveys \& Tutorials},
  vol.~21, no.~3, pp. 2568--2599, 2019.

\bibitem{jia2025optimum}
M.~Jia, J.~Chen, Y.-C. Liang, and P.-Y. Kam, ``Optimum noncoherent detection of
  constant-envelope signals using received signal magnitudes--energy detection
  and amplitude detection,'' \emph{arXiv preprint arXiv:2502.17897}, 2025.

\bibitem{wang2020tcom}
T.~Wang, H.~Xiong, H.~Ding, and L.~Zheng, ``Tdoa-based joint synchronization
  and localization algorithm for asynchronous wireless sensor networks,''
  \emph{IEEE Transactions on Communications}, vol.~68, no.~5, pp. 3107--3124,
  2020.

\bibitem{MIMOradar}
A.~M. {Haimovich}, R.~S. {Blum}, and L.~J. {Cimini}, ``Mimo radar with widely
  separated antennas,'' \emph{IEEE Signal Processing Magazine}, vol.~25, no.~1,
  pp. 116--129, 2008.

\bibitem{lth2021vtc}
T.~Liang, J.~Yang, and T.~Zhang, ``Uav aided vehicle positioning with imperfect
  data association,'' in \emph{2021 IEEE 93rd Vehicular Technology Conference
  (VTC2021-Spring)}, 2021, pp. 1--6.

\bibitem{lth2023vtc}
T.~Liang and T.~Zhang, ``Simultaneous localization and tracking for
  uav-enhanced positioning network,'' in \emph{2023 IEEE 98th Vehicular
  Technology Conference (VTC2023-Fall)}, 2023, pp. 1--5.

\bibitem{yjy2021vtc}
J.~Yang, T.~Liang, and T.~Zhang, ``Deployment optimization in uav aided vehicle
  localization,'' in \emph{2021 IEEE 93rd Vehicular Technology Conference
  (VTC2021-Spring)}, 2021, pp. 1--6.

\bibitem{clutterproperty}
H.~Zhu, Z.~Zhu, and F.~Su, ``Clutter properties and suppression methods of
  hyper sonic airborne radar,'' in \emph{2018 14th IEEE International
  Conference on Signal Processing (ICSP)}, 2018, pp. 859--862.

\bibitem{MTIbasic}
M.~Ash, M.~Ritchie, and K.~Chetty, ``On the application of digital moving
  target indication techniques to short-range {FMCW} radar data,'' \emph{IEEE
  Sensors Journal}, vol.~18, no.~10, pp. 4167--4175, 2018.

\bibitem{ezu2019radar}
M.~Ezuma, O.~Ozdemir, C.~K. Anjinappa, W.~A. Gulzar, and I.~Guvenc, ``Micro-uav
  detection with a low-grazing angle millimeter wave radar,'' in \emph{2019
  IEEE Radio and Wireless Symposium (RWS)}, 2019, pp. 1--4.

\bibitem{kam2018taes}
A.~Kammoun, R.~Couillet, F.~Pascal, and M.-S. Alouini, ``Optimal design of the
  adaptive normalized matched filter detector using regularized tyler
  estimators,'' \emph{IEEE Transactions on Aerospace and Electronic Systems},
  vol.~54, no.~2, pp. 755--769, 2018.

\bibitem{rahman2018radar}
S.~Rahman and D.~A. Robertson, ``Radar micro-doppler signatures of drones and
  birds at k-band and w-band,'' \emph{Scientific reports}, vol.~8, no.~1, p.
  17396, 2018.

\bibitem{RadarConfMicro}
M.~Jian, Z.~Lu, and V.~C. Chen, ``Experimental study on radar micro-doppler
  signatures of unmanned aerial vehicles,'' in \emph{2017 IEEE Radar Conference
  (RadarConf)}.\hskip 1em plus 0.5em minus 0.4em\relax IEEE, 2017, pp.
  0854--0857.

\bibitem{fu2018cm}
H.~Fu, S.~Abeywickrama, L.~Zhang, and C.~Yuen, ``Low-complexity portable
  passive drone surveillance via sdr-based signal processing,'' \emph{IEEE
  Communications Magazine}, vol.~56, no.~4, pp. 112--118, 2018.

\bibitem{JiaShaKer:C20}
M.~Jia, S.~Li, J.~L. Kernec, S.~Yang, F.~Fioranelli, and O.~Romain, ``Human
  activity classification with radar signal processing and machine learning,''
  in \emph{2020 International Conference on UK-China Emerging Technologies
  (UCET)}, 2020, pp. 1--5.

\bibitem{vov2025tap}
D.~Vovchuk, M.~Khobzei, V.~Tkach, O.~Eliiashiv, O.~Tzidki, K.~Grotov, A.~Glam,
  and P.~Ginzburg, ``Micro-doppler-coded drone identification via resonant
  tagging,'' \emph{IEEE Transactions on Antennas and Propagation}, vol.~73,
  no.~6, pp. 3917--3927, 2025.

\bibitem{EMD1}
A.~{Brewster} and A.~{Balleri}, ``Extraction and analysis of micro-doppler
  signatures by the empirical mode decomposition,'' in \emph{2015 IEEE Radar
  Conference (RadarCon)}, May 2015, pp. 0947--0951.

\bibitem{yu2025tifs}
N.~Yu, J.~Wu, C.~Zhou, Z.~Shi, and J.~Chen, ``Open set learning for rf-based
  drone recognition via signal semantics,'' \emph{IEEE Transactions on
  Information Forensics and Security}, vol.~19, pp. 9894--9909, 2024.

\bibitem{khan2024tvt}
M.~U. Khan, M.~Dil, M.~Z. Alam, F.~A. Orakazi, A.~M. Almasoud, Z.~Kaleem, and
  C.~Yuen, ``Safespace mfnet: Precise and efficient multifeature drone
  detection network,'' \emph{IEEE Transactions on Vehicular Technology},
  vol.~73, no.~3, pp. 3106--3118, 2024.

\bibitem{Liu2024ics}
F.~Liu, X.~Sun, H.~Zhang, Y.~Wang, X.~Qu, and X.~Yang, ``Multipath ghost
  suppression based on frequency domain filtering for through-the-wall radar,''
  in \emph{2024 IEEE International Conference on Signal, Information and Data
  Processing (ICSIDP)}, 2024, pp. 1--6.

\bibitem{zhou2021vtc}
Y.~Zhou, X.~Huang, D.~Zhang, B.~Li, J.~Zeng, and F.~Wu, ``5g-based measurements
  and characterizations of low-altitude tethered balloon multipath channel,''
  in \emph{2021 IEEE 93rd Vehicular Technology Conference (VTC2021-Spring)},
  2021, pp. 1--5.

\bibitem{ghostbackground}
J.~{Liu}, L.~{Kong}, X.~{Yang}, and Q.~H. {Liu}, ``First-order multipath
  ghosts' characteristics and suppression in mimo through-wall imaging,''
  \emph{IEEE Geoscience and Remote Sensing Letters}, vol.~13, no.~9, pp.
  1315--1319, 2016.

\bibitem{beamintersection}
I.~Trofymov, V.~Tiutiunnyk, A.~Dudush, A.~Shevchenko, and I.~Medinets,
  ``Deghosting method for multistatic radar systems with cooperative
  receiving,'' in \emph{2020 IEEE Ukrainian Microwave Week (UkrMW)}, 2020, pp.
  1--4.

\bibitem{bistaticghost}
A.~T. Abdalla and A.~H. Muqaibel, ``Single-view bistatic sparse reconstruction
  in twri exploiting ghost's aspect dependence feature,'' in \emph{2016 IEEE
  Wireless Communications and Networking Conference}, 2016, pp. 1--5.

\bibitem{ghostfusion}
H.~{Li}, G.~{Cui}, S.~{Guo}, L.~{Kong}, and X.~{Yang}, ``Target tracking and
  ghost mitigation based on multi-view through-the-wall radar imaging,'' in
  \emph{2020 IEEE Radar Conference (RadarConf20)}, 2020, pp. 1--5.

\bibitem{ghostrangedoppler}
R.~Feng, E.~De~Greef, M.~Rykunov, H.~Sahli, S.~Pollin, and A.~Bourdoux,
  ``Multipath ghost recognition for indoor mimo radar,'' \emph{IEEE
  Transactions on Geoscience and Remote Sensing}, pp. 1--10, 2021.

\bibitem{machinelearningghost}
F.~{Kraus}, N.~{Scheiner}, W.~{Ritter}, and K.~{Dietmayer}, ``Using machine
  learning to detect ghost images in automotive radar,'' in \emph{2020 IEEE
  23rd International Conference on Intelligent Transportation Systems (ITSC)},
  2020, pp. 1--7.

\bibitem{Junyang}
J.~{Shen} and A.~F. {Molisch}, ``Estimating multiple target locations in
  multi-path environments,'' \emph{IEEE Transactions on Wireless
  Communications}, vol.~13, no.~8, pp. 4547--4559, 2014.

\bibitem{NETWORKLOCALIZATION}
M.~Z. Win, Y.~Shen, and W.~Dai, ``A theoretical foundation of network
  localization and navigation,'' \emph{Proceedings of the IEEE}, vol. 106,
  no.~7, pp. 1136--1165, 2018.

\bibitem{lth2024wcl}
T.~Liang, T.~Zhang, B.~Cao, and Q.~Zhang, ``Sensing, communication, and control
  co-design for energy-efficient {UAV}-aided data collection,'' \emph{IEEE
  Wireless Communications Letters}, vol.~13, no.~10, pp. 2852--2856, 2024.

\bibitem{JASP}
T.~Zhang, C.~Qin, A.~F. Molisch, and Q.~Zhang, ``Joint allocation of spectral
  and power resources for non-cooperative wireless localization networks,''
  \emph{IEEE Transactions on Communications}, vol.~64, no.~9, pp. 3733--3745,
  2016.

\bibitem{lth2024wcnc}
T.~Liang, Z.~Yu, T.~Zhang, S.~Zhou, W.~Liu, D.~Li, and Z.~Niu, ``Joint frame
  structure and beamwidth optimization for integrated localization and
  communication,'' in \emph{2024 IEEE Wireless Communications and Networking
  Conference (WCNC)}, 2024, pp. 1--6.

\bibitem{RLallocation}
B.~Peng, G.~Seco-Granados, E.~Steinmetz, M.~Fr?hle, and H.~Wymeersch,
  ``Decentralized scheduling for cooperative localization with deep
  reinforcement learning,'' \emph{IEEE Transactions on Vehicular Technology},
  vol.~68, no.~5, pp. 4295--4305, 2019.

\bibitem{yang2021vtc}
C.~Yang, F.~Liu, and T.~Zhang, ``{MSE} based resource optimization in wireless
  localization networks,'' in \emph{2021 IEEE 93rd Vehicular Technology
  Conference (VTC2021-Spring)}, 2021, pp. 1--6.

\bibitem{Richards2014Fundamentals}
M.~A. Richards, \emph{Fundamentals of radar signal processing,2nd ed}, 2014.

\bibitem{richards2005fundamentals}
------, \emph{Fundamentals of radar signal processing}.\hskip 1em plus 0.5em
  minus 0.4em\relax Tata McGraw-Hill Education, 2005.

\bibitem{Music2016algorithm}
L.~{Liu}, J.~{Xu}, G.~{Wang}, X.~{Xia}, Y.~{Gao}, and T.~{Long}, ``An extended
  dimension music method for doa estimation of multiple real-valued sources,''
  in \emph{2016 CIE International Conference on Radar (RADAR)}, Oct 2016, pp.
  1--5.

\bibitem{chen2006micro}
V.~C. Chen, F.~Li, S.-S. Ho, and H.~Wechsler, ``Micro-doppler effect in radar:
  phenomenon, model, and simulation study,'' \emph{IEEE Transactions on
  Aerospace and electronic systems}, vol.~42, no.~1, pp. 2--21, 2006.

\bibitem{Micro2018}
P.~{Sathe}, A.~{Dyana}, K.~P. {Ray}, D.~{Shashikiran}, and A.~{Vengadarajan},
  ``Helicopter main and tail rotor blade parameter extraction using
  micro-doppler,'' in \emph{2018 19th International Radar Symposium (IRS)},
  June 2018, pp. 1--10.

\bibitem{Chen2011The}
V.~C. Chen and I.~Ebrary, ``The micro-doppler effect in radar,'' 2011.

\bibitem{lthiotj}
T.~Liang, T.~Zhang, J.~Yang, D.~Feng, and Q.~Zhang, ``{UAV}-aided positioning
  systems for ground devices: Fundamental limits and algorithms,'' \emph{IEEE
  Internet of Things Journal}, vol.~9, no.~15, pp. 13\,470--13\,485, 2022.

\bibitem{shen2010fundamental}
Y.~Shen and M.~Z. Win, ``Fundamental limits of wideband localization part i: A
  general framework,'' \emph{IEEE Transactions on Information Theory}, vol.~56,
  no.~10, pp. 4956--4980, 2010.

\end{thebibliography}

\end{document}